\documentclass[aps,pra,11pt,amsmath,notitlepage,tightenlines,superscriptaddress,eqsecnum,floatfix,showpacs,longbibliography]{revtex4-1}

\usepackage{bm,hyperref,ifpdf,tikz}
\usepackage[utf8]{inputenc} 

\newcommand{\tr}{{\rm tr}}
\newcommand{\cM}{\mathcal{M}}
\newcommand{\cQ}{\mathcal{Q}}
\newcommand{\cC}{\mathcal{C}}
\newcommand{\sA}{\mathcal{A}}
\newcommand{\sB}{\mathcal{B}}
\newcommand{\sD}{\mathcal{D}}

\begin{document}

\ifpdf
\DeclareGraphicsExtensions{.pdf, .jpg, .tif}
\else
\DeclareGraphicsExtensions{.png, .jpg}
\fi

\title{Entropic measures of nonclassical correlations}

\author{Matthias D.~Lang}
\affiliation{Center for Quantum Information and Control, University of New Mexico, Albuquerque, New Mexico 87131-0001, USA}

\author{Carlton M.~Caves}
\affiliation{Center for Quantum Information and Control, University of New Mexico, Albuquerque, New Mexico 87131-0001, USA}
\affiliation{Centre for Engineered Quantum Systems, School of Mathematics and Physics, University of Queensland, Brisbane, Queensland 4072, Australia}

\author{Anil Shaji}
\email{shaji@iisertvm.ac.in}
\affiliation{School of Physics, Indian Institute of Science Education and Research\,--\,Thiruvananthapuram, CET Campus, Kerala, India 695016}

\begin{abstract}
A framework for categorizing entropic measures of nonclassical correlations in bipartite quantum states is presented.  The measures are based on the difference between a quantum entropic quantity and the corresponding classical quantity obtained from measurements on the two systems.  Three types of entropic quantities are used, and three different measurement strategies are applied to these quantities.  Many of the resulting measures of nonclassical correlations have been proposed previously.  Properties of the various measures are explored, and results of evaluating the measures for two-qubit quantum states are presented.
\end{abstract}

\date{\today}

\maketitle

\section{Introduction}
\label{sec:intro}

Maxwell demons observe a physical system and use the information obtained to extract work from the system~\cite{maxwell1891}.  For multipartite systems, we can distinguish quantum Maxwell demons, which have knowledge of the entire density operator and can manipulate and make measurements on the joint system, from classical demons, which can only perform operations and make measurements on the subsystems of the multipartite system.  Because a single classical demon cannot be everywhere at the same time, it must recruit local demons to gather, process, and use information about the local systems; thus it is better to think of a classical demon as a collection of local demons.  These local demons might or might not be allowed to communicate with each other using classical channels.  The amount of work that the two kinds of demons, quantum and classical, can extract from a given multipartite quantum state by employing protocols within each demon's means is a way of comparing quantum-information-processing protocols with classical ones.

This demonology~\cite{zurek00a,Oppenheim2002a,zurek03b,brodutch10a} is but one of several attempts~\cite{zurek00a,ollivier01a,henderson01a,rajagopal02a,zurek03b,Piani2008a,luo08a,wu09a,brodutch10a,modi10a} to track down and quantify the correlations that exist in multipartite quantum states.  The nonclassical part of these correlations is not just quantum entanglement, even though entanglement is a part of it.  The open question of pinning down why mixed-state quantum algorithms can solve certain problems exponentially faster than the best known classical ones~\cite{jozsa03a}, even in the absence of any significant entanglement, is one of the main motivations behind studying the nonclassical correlations in quantum states other than entanglement~\cite{datta07a,datta08a,datta08b,datta09a,Eastin2010au}.

We consider only bipartite states in this paper.  For our numerical work, the discussion is specialized yet further to states of two qubits.  Correlations between systems can be quantified in terms of correlation coefficients and covariance matrices or in terms of entropic measures like mutual information.  We choose the latter approach as the preferred one in information theory.  The aim of the paper is to formulate a framework in terms of which the several entropy-based measures of nonclassical correlations that have been proposed can be classified and understood.  Constructing the framework leads to two new measures we have not seen previously in the literature.  The focus here is not so much on unifying various measures, as in Ref.~\cite{modi10a}, but rather on clarifying the relationships among them.

The setting for our framework is two systems, $A$ and $B$, with a joint quantum state $\rho_{AB}$. We consider three types of nonclassical-correlation measures, $\cM(\rho_{AB})$, between $A$ and $B$:
\begin{enumerate}
	\item Mutual-information-based measures.
	\item Conditional-entropy-based measures.
	\item Demon-based (joint-entropy-based) measures.
\end{enumerate}
The type-2 correlation measures can be asymmetric between $A$ and $B$ because conditional entropy is typically asymmetric.

As Landauer pointed out, when talking about demons, erasure of the demon's memory---and the associated thermodynamic cost---is an essential feature for assessing what a demon can do~\cite{landauer61a}.  As we mentioned above, a classical demon that works on a bipartite quantum system is best thought of as two local demons working in concert.  Whether the two demons can communicate impacts their ability to co\"operate. So the demon-based measures are thus further divided into two classes:
\begin{enumerate}
	\item[i.] Erasure without communication between the demons.
	\item[ii.] Erasure with communication between the demons.
\end{enumerate}

All the measures of nonclassical correlations we consider here are constructed as the difference between a quantum entropic measure, $\cQ(\rho_{AB})$, and its classical counterpart, $\cC(\rho_{AB})$, which is derived from the probabilities for results of local measurements on one or both of the subsystems.  The thinking behind this construction is that $\cQ$ quantifies some notion of all the correlations in the system, whereas the corresponding classical $\cC$ captures only the corresponding classical correlations.  The difference, $\cM=\cQ-\cC$, is therefore a way of quantifying the nonclassical correlations in the quantum state.

The results of local measurements are all that local classical observers (demons) can access, and these measurement results are used to probe the correlations (if any) between $A$ and $B$.  We do not want, however, our measure of nonclassical correlations to depend on the specifics of the measurement performed.  Hence, in its construction, the classical measure, $\cC(\rho_{AB}$), is maximized over all possible measurements within specific measurement strategies that are defined beforehand.  In some instances, when maximization is necessary, we are able to show that the maximum is attained on rank-one POVMs; in other cases, we restrict the maximization to rank-one POVMs.  We give a full discussion of these different situations and the issues surrounding rank-one POVMs after we have developed our framework.

We thus imagine that there are classical observers $A$ and $B$---demons or otherwise---who have access to the two parts of the bipartite system.  We allow these observers to employ one of three measurement strategies:
\begin{enumerate}
	\item[a.] Local, rank-one-projector measurements in the eigenbases of the marginal density operators.
	\item[b.] Unconditioned local measurements.
	\item[c.] Conditioned local measurements.
\end{enumerate}
For strategy~(a), the local measurements are unique modulo degeneracies in the marginal density operators.  The other two strategies require maximization of the classical measure $\cC$ over the measurements allowed by the strategy.  The first two measurement strategies do not require the observers to communicate with each other, but the last one does.  Consequently, the first two strategies are symmetric between $A$ and $B$.  For the third strategy, $A$ performs a measurement and communicates the result to $B$, who can then condition his measurement on the result communicated by $A$.  This makes the nonclassical correlation measures that are based on the third measurement strategy asymmetric between $A$ and $B$.

We now have three types of correlation measures and three measurement strategies, and we can label the resulting correlation measures with the type and the strategy.  For example, $\cM_{1b}$ refers to the nonclassical correlation measure constructed as the difference between quantum and classical mutual informations, where unconditioned local measurements are used to construct the classical mutual information.

There is a natural hierarchy in the three types of measurements strategies.  Allowing arbitrary, unconditioned local measurements, as in strategy~(b), is a restriction of the conditioned local measurements
of strategy~(c), since to get~(b) from~(c), observer $B$ simply chooses to ignore any communication $A$ might have sent regarding her measurement results.  Likewise, measuring in the local eigenbases of the marginal density operators, as in strategy~(a), is a restriction of the arbitrary, unconditioned local measurements of strategy~(b).  Thus, when we maximize over the measurements in a particular strategy, the classical measure $\cC$ cannot decrease---and generally it increases---as we move from~(a) to~(b) to~(c).  This is saying that the more general the measurements the local observers are allowed to do, the more they can expect to discover about any classical correlations that exist between the subsystems.  Since our nonclassical-correlation measure $\cM$ is the difference between $\cQ$ and $\cC$, $\cM$ cannot increase---and generally it decreases---as we move from~(a) to~(b) to~(c), i.e., $\cM_{ja}\ge\cM_{jb}\ge\cM_{jc}$ for $j=1,2,3$.

In Sec.~\ref{sec:framework} we formulate our framework: Sec.~\ref{sec:measures} reviews the bipartite entropic information measures that we use in constructing our framework; Sec.~\ref{sec:localmeasurements} spells out the description of local measurements for strategies~(a)--(c); Sec.~\ref{sec:Measures} defines the nonclassical-correlation measures and discusses relations among them; and Sec.~\ref{sec:POVMs} considers the issues raised by assuming the local measurements are described by rank-one POVMs and also whether one can specialize further to measurements described by rank-one projectors.  In Sec.~\ref{sec:numerics} we present numerical results comparing the various  measures for two-qubit states, assuming that the local measurements can be described by orthogonal rank-one projection operators.  A concluding Sec.~\ref{sec:conclusion} draws attention to outstanding questions, and several appendices provide additional information.

\section{Framework for entropic measures of nonclassical correlations}
\label{sec:framework}

In this section we develop our framework for measures of nonclassical correlations and explore properties of the various measures the framework leads to.

\subsection{Entropic measures of information and correlation}
\label{sec:measures}

\begin{figure}[!htb]
\begin{center}
\includegraphics[width=10cm]{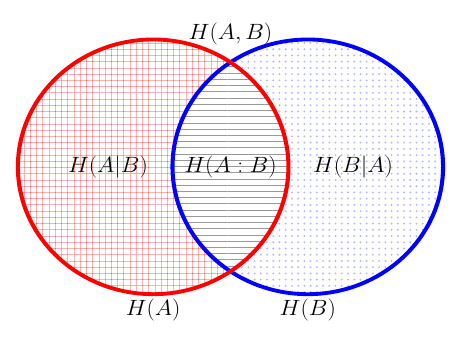}\\
\end{center}
\caption{(Color online) The (red) circle on the left denotes the entropy associated with system $A$; the (blue) circle on the right denotes the entropy associated with system $B$.  The area on the right filled in with (blue) dots is the information missing about $B$ given complete information about $A$; this area denotes the conditional entropy $H(B|A)$.  Similarly, the area on the left filled in with the (red) grid denotes $H(A|B)$.  The overlap between the two circles, filled with horizontal lines, denotes the mutual information $H(A:B)$, which is the information contained in $A$ about $B$ and vice versa.  The combined envelope of the two circles is the joint entropy $H(A,B)$.  From the diagram, we have
$H(B|A)=H(A,B)-H(A)=H(B)-H(A:B)$ and $H(A|B)=H(A,B)-H(B)=H(A)-H(A:B)$.  For a classical joint probability distribution, the entropic measures are all Shannon entropies or relative Shannon entropies---thus they are guaranteed to be nonnegative---and they are related as the diagram depicts.  For a bipartite quantum state, the joint quantum von Neumann entropy, $S(A,B)$, and the marginal von Neumann entropies, $S(A)$ and $S(B)$, replace $H(A,B)$, $H(A)$, and $H(B)$.  The measures are related as depicted in the diagram, because the quantum conditional entropies, $S(B|A)$ and $S(A|B)$, and the quantum mutual information, $S(A:B)$, are \textit{defined\/} by these relations.  The difference is that $S(B|A)$ and $S(A|B)$, as so defined, can be negative, and thus the quantum mutual information $S(A:B)$ can be bigger than the marginal entropies, $S(A)$ and $S(B)$, and bigger than the joint entropy $S(A,B)$.\label{fig:Venn}}
\end{figure}

Entropic measures of information quantify how much information can be extracted from a system or, more poetically, how much information is ``missing'' about the fine-grained state of the system.  Figure~\ref{fig:Venn} is a useful pictorial representation of the relationships among the entropies and entropic measures of correlation that apply to bipartite systems.  The figure provides an accurate representation for classical entropies.  In the quantum case, some of the quantities cannot be represented or are misrepresented by this diagram, but even so, the diagram is a useful tool because it captures correctly the relationships among the various entropies.

For a bipartite state $\rho_{AB}$ of systems $A$ and $B$, the quantum entropic quantities that will be used in the ensuing discussion are the following:
\begin{enumerate}
	\item $S(A, B)=S(\rho_{AB})=-\tr(\rho_{AB}\log\rho_{AB})$, the joint von Neumann entropy of the whole system.
	\item $S(A)=S(\rho_A)=-\tr_A(\rho_A\log\rho_A)$ and $S(B)=S(\rho_B)=-\tr_B(\rho_B\log\rho_B)$, the von Neumann entropies of the marginal density operators.
	\item $S(B|A)=S(A,B)-S(A)$ and $S(A|B) = S(A,B) - S(B)$, the quantum conditional entropies.
	\item $S(A:B) = S(A) + S(B) - S(A,B)$, the quantum mutual information, which is related to the quantum conditional entropies by $S(A:B) = S(B)- S(B|A)=S(A)-S(A|B)$.  The quantum mutual information can also be written as a quantum relative entropy,
\begin{equation}\label{eq:qmutualrelativeentropy}
S(A:B)=S(\rho_{AB}||\rho_A\otimes\rho_B)\;,
\end{equation}
where the relative entropy is defined by
\begin{equation}\label{eq:qrelativeentropy}
S(\rho||\sigma)=-S(\rho)-\tr(\rho\log\sigma)\;.
\end{equation}
\end{enumerate}

Local measurements on the bipartite quantum system are described by a joint probability distribution $p_{ab}$ for outcomes labeled by $a$ and $b$.  Bayes's theorem relates the joint, conditional, and marginal distributions: $p_{b|a}p_a=p_{ab}=p_{a|b}p_b$.  These distributions are used to define the classical information measures:
\begin{enumerate}
	\item $H(A, B)=H(p_{ab})=-\sum_{a,b}p_{ab}\log p_{ab}$, the Shannon entropy of the joint distribution~$p_{ab}$.
	\item $H(A)=H(p_a)=-\sum_a p_a\log p_a$ and $H(B)=H(p_b)=-\sum_a p_b\log p_b$, the Shannon entropies of the marginal distributions, $p_a$ and $p_b$.
	\item $H(B|A)\!=\!H(A,B)-H(A)=\sum_ap_aH(B|a)$ and $H(A|B)\!=\!H(A,B)-H(B)=\sum_a p_bH(A|b)$, the classical conditional entropies.  $H(B|a)\!=\!-\sum_bp_{b|a}\log p_{b|a}$ and $H(A|b)\!=\!-\sum_ap_{a|b}\log p_{a|b}$ are the Shannon entropies of the conditional distributions $p_{b|a}$ and $p_{a|b}$; the conditional entropies are averages of $H(B|a)$ over $p_a$ and $H(A|b)$ over~$p_b$.
	\item $H(A:B) = H(A) + H(B) - H(A,B)=\sum_{a,b}p_{ab}\log(p_{ab}/p_ap_b)$, the classical mutual information.  $H(A:B)$ is the relative information of the joint distribution $p_{ab}$ with respect to the product of the marginals, $p_ap_b$,
\begin{equation}\label{eq:cmutualrelativeentropy}
H(A:B)=H(p_{ab}||p_ap_b)\;;
\end{equation}
the classical relative information, which is always nonnegative, is defined by
\begin{equation}\label{eq:crelativeentropy}
H(p_j||q_j)=\sum_jp_j\log(p_j/q_j)
=-H(p_j)-\sum_jp_j\log q_j\;.
\end{equation}
We also have $H(A:B)=H(B)-H(B|A)=H(A)-H(A|B)$.
\end{enumerate}

Figure~\ref{fig:Venn} summarizes the relations among the classical entropies; it works because the classical conditional entropies and the classical mutual information are all nonnegative.  This leads to several
inequalities that can be read off Fig.~\ref{fig:Venn}.  For example,
we can see that
\begin{equation}
	\label{eq:ineq1}
	\max\!\big( H(A),H(B)\big) \leq H(A,B) \leq H(A) + H(B)\;.
\end{equation}
The lower bound on $H(A,B)$ is saturated when knowing one subsystem completely determines the other (the two circles in Fig.~\ref{fig:Venn} are either identical or become nested), i.e., $H(A:B)=\min\big( H(A),H(B)\big)$.  The upper bound is saturated when there are no correlations between $A$ and $B$, i.e., $H(A:B)=0$, so determining one subsystem gives no information about the other (the two circles in Fig.~\ref{fig:Venn} are disjoint).  For quantum entropies the lower bound in Eq.~(\ref{eq:ineq1}) does not hold, which is equivalent to saying the quantum conditional entropies can be negative.  The simplest counter-example is a two-qubit Bell state: the joint state is pure and, hence, has zero entropy, but the marginal states are completely mixed, so their entropies are maximal and both equal to one.

\subsection{Local measurements}
\label{sec:localmeasurements}

We now spell out the general description of the local measurements that applies to measurement strategies~(a)--(c).  Although we only need measurement statistics---and, hence, only need POVMs---to evaluate the classical entropic measures, we start our description with quantum operations, partly to be general and partly so we can deal with post-measurement states in a subsequent discussion of Maxwell demons.

The measurement on $A$ is described by quantum operations~\cite{nielsen00a} that are labeled by the possible outcomes $a$ of the measurement on $A$:
\begin{equation}
\sA_a=\sum_\alpha A_{a\alpha}\odot A_{a\alpha}^\dagger\;.
\end{equation}
The quantum operation is applied to a density operator by inserting the density operator in place of the $\odot$.  The operators $A_{a\alpha}$, the Kraus operators of $\sA_a$, combine to give the POVM element for outcome~$a$,
\begin{equation}
E_a=\sum_\alpha A_{a\alpha}^\dagger A_{a\alpha}\;,
\end{equation}
and the POVM elements satisfy a completeness relation, $I_A=\sum_a E_a$.

The absence of communication in strategies~(a) and~(b) makes them quite straightforward.  The measurement on $B$ is described by a set of quantum operations,
\begin{equation}
\sB_b=\sum_\beta B_{b\beta}\odot B_{b\beta}^\dagger\;.
\end{equation}
These give POVM elements
\begin{equation}
F_b=\sum_\beta B_{b\beta}^\dagger B_{b\beta}\;,
\end{equation}
which satisfy a completeness relation $I_B=\sum_b F_b$.  The state of the joint system after measurements with outcomes~$a$ and~$b$ is $\rho_{AB|ab}=\sA_a\otimes\sB_b(\rho_{AB})/p_{ab}$, where
\begin{equation}
p_{ab}=\tr\bigl(\sA_a\otimes\sB_b(\rho_{AB})\bigr) = \tr(E_a\otimes F_b\rho_{AB})
\end{equation}
is the joint probability for outcomes $a$ and $b$.  The post-measurement joint state and the joint probability marginalize to the subsystems in the standard way.

We need to be more careful with strategy~(c) because of the communication from~$A$ to~$B$.  We handle strategy~(c) in a general way that allows us to interpolate between~(b) and the extreme case of~(c) in which every outcome $a$ leads to a different measurement on $B$.  We do this by introducing a set~$C$ whose elements~$c$ label the possible measurements to be made on~$B$.  We let~$A$ stand for the set of outcomes~$a$, and we define a function $c(a)$ that maps an outcome~$a$ to the corresponding value in $C$.  We let $A_c=\{a\mid c(a)=c\}$ be the subset of $A$ that leads to the $B$ measurement labeled by $c$.  The subsets $A_c$ partition $A$ into disjoint subsets.  We can regard $C$ as another variable in our analysis; it is a coarse graining of the measurement on $A$.  Formally, we have that $C$ is perfectly correlated with $A$, i.e., $p_{c|a}=\delta_{c,c(a)}$, implying that $H(C|A)=0$ and $H(A:C)=H(C)$.  Should there be only one possible measurement on $B$, i.e., only one value of $c$, then there is no communication, and the situation reduces to strategy~(b).  The extreme case of~(c) corresponds to having a different value of $c$ for each outcome $a$, in which case there is no difference between the outcome set~$A$ and the set~$C$.

The state of the joint system after the measurement on $A$ yields outcome~$a$ is $\rho_{AB|a}=\sA_a(\rho_{AB})/p_a$, where
\begin{equation}
p_a=\tr\bigl(\sA_a(\rho_{AB})\bigr)=\tr_A(E_a\rho_A)
\end{equation}
is the probability for outcome~$a$.  The state of system~$B$, conditioned on outcome~$a$, is
\begin{equation}\label{eq:rhoBa}
\rho_{B|a}=\tr_A(\rho_{AB|a})=\frac{\tr_{A}\bigl(E_a\rho_{AB}\bigr)}{p_a}\;;
\end{equation}
notice that this is determined by the POVM element $E_a$.  The probability for making measurement~$c$ on $B$ follows formally from
\begin{equation}
p_c=\sum_a p_{c|a}p_a=\sum_{a\in A_c}p_a=\tr_A(E_c\rho_A)\;.
\end{equation}
Here we introduce coarse-grained POVM elements for the measurement on $A$, labeled by the measurement to be made on $B$:
\begin{equation}
E_c=\sum_{a\in A_c}E_a\;.
\end{equation}
Notice that if there is only one possible measurement on $B$, i.e., only one value of $c$, then $E_c=I_A$; when there is a different measurement for each outcome~$a$, the POVM elements $E_c$ are the same as the POVM elements~$E_a$.  We also have the state of $B$ conditioned on the coarse-grained outcome~$c$:
\begin{equation}\label{eq:rhoBc}
\rho_{B|c}=\frac{\tr_{A}\bigl(E_c\rho_{AB}\bigr)}{p_c}\;.
\end{equation}
Notice that Eqs.~(\ref{eq:rhoBa}) and (\ref{eq:rhoBc}) imply that
\begin{equation}\label{eq:rhoBensembles}
\rho_B=\sum_ap_a\rho_{B|a}=\sum_cp_c\rho_{B|c}\;.
\end{equation}

We turn our attention now to the measurements on $B$.  We let $B$ stand for the set of all outcomes on $B$ for all the possible measurements on $B$.  We define a function $c(b)$ that maps an outcome $b$ to the measurement~$c$ in which it occurs, and we define $B_c=\{b\mid c(b)=c\}$ to be the subset of $B$ outcomes for the measurement labeled by~$c$.  The subsets $B_c$ partition the set of all possible outcomes on $B$ into disjoint subsets.  We again have perfect correlation, i.e.,
$p_{c|b}=\delta_{c,c(b)}$, implying that $H(C|B)=0$ and $H(B:C)=H(C)$.

The measurement on $B$ that is labeled by $c$ is described by quantum operations
\begin{equation}
\sB_{b|c}=\sum_\beta B_{b\beta|c}\odot B_{b\beta|c}^\dagger\;,
\end{equation}
The Kraus operators give the POVM elements for this measurement,
\begin{equation}
F_{b|c}=\sum_\beta B_{b\beta|c}^\dagger B_{b\beta|c}\;,
\end{equation}
and these satisfy a completeness relation $I_B=\sum_{b\in B_c} F_{b|c}$.  In sums over $b$, we can let the sum run over the outcomes of all the possible measurements on $B$ by the artifice of defining $B_{b\beta|c}=0$ for $b\notin B_c$ and, hence, $F_{b|c}=0$ for $b\notin B_c$.

The state of the joint system, conditioned on outcomes $a$ and $b$, is
\begin{equation}\label{eq:rhoABgivenab}
\rho_{AB|ab}=\frac{\sA_a\otimes\sB_{b|c(a)}(\rho_{AB})}{p_{ab}}= \frac{\sB_{b|c(a)}(\rho_{AB|a})}{p_{b|a}}\;,
\end{equation}
where
\begin{equation}\label{eq:pab}
p_{ab}=\tr\bigl(\sA_a\otimes\sB_{b|c(a)}(\rho_{AB})\bigr)
=\tr(E_a\otimes F_{b|c(a)}\rho_{AB})=p_a\tr_B(F_{b|c(a)}\rho_{B|a})
\end{equation}
is the joint probability for $a$ and $b$ and
\begin{equation}\label{eq:pbgivena}
p_{b|a}=\tr\bigl(\sB_{b|c(a)}(\rho_{AB|a})\bigr)
=\tr(F_{b|c(a)}\rho_{B|a})
\end{equation}
is the conditional probability for $b$ given $a$.  Notice that $p_{ab}$ and $p_{b|a}$ are nonzero only if $b\in B_{c(a)}$ or, equivalently, only if $a\in A_{c(b)}$.

For our purposes, it is easier to work with the coarse-grained outcomes $c$, which specify the measurements on $B$.  Indeed, the joint probability for~$b$ and $c$~is
\begin{equation}
p_{bc}=\sum_a p_{c|ab}p_{ab}=\sum_{a\in A_c}p_{ab}=
\tr(E_c\otimes F_{b|c}\rho_{AB})=
p_c\tr_B(F_{b|c}\rho_{B|c})\;.
\end{equation}
Notice that $p_{bc}$ is nonzero only if $b\in B_c$.  Thus the conditional probability of $b$ given $c$ takes the form
\begin{equation}\label{eq:pbgivenc}
p_{b|c}=\frac{p_{bc}}{p_c}=\tr_B(F_{b|c}\rho_{B|c})\;,
\end{equation}
and the unconditioned probability for $b$ is
\begin{equation}
p_b=\sum_c p_{bc}=\tr(E_{c(b)}\otimes F_{b|c(b)}\rho_{AB})
=p_{c(b)}\tr_B(F_{b|c(b)}\rho_{B|c(b)})\;.
\end{equation}

\subsection{Measures of nonclassical correlations}
\label{sec:Measures}

In this subsection we formulate our framework for entropic measures of nonclassical correlations, considering in turn the three types of
measures introduced in Sec.~\ref{sec:intro} and for each type, the three local measurement strategies,~(a), (b), and~(c).  For strategy~(a), the local measurements are in the eigenbases of the marginal density operators.  \textit{For strategies~(b) and~(c), we assume that the measurements are described by rank-one POVMs}, which means that $E_a$ and $F_{b|c}$ are multiples of rank-one projection operators.  We discuss this assumption in Sec.~\ref{sec:POVMs}.

To compare and relate the various measures, we rely on two inequalities that relate the quantum and the classical entropies: the \textit{POVM inequality\/} (see Appendix~\ref{app:POVMinequality} for a proof) and the \textit{ensemble inequality\/}~\cite{nielsen00a}.

The POVM inequality relates the quantum entropy for a state $\rho$ to the classical entropy for probabilities $p_j=\tr(E_j\rho)$ obtained from (nonzero) POVM elements $E_j$:
\begin{equation}\label{eq:povm1}
	H(p_j)+\sum_j p_j\log(\tr E_j)
    =-\sum_j p_j\log\!\left(\frac{p_j}{\tr E_j}\right)
    \ge S(\rho)\;.
\end{equation}
A rank-one POVM is one such that all the POVM elements are rank-one, i.e., $E_j=\mu_j P_j$, where $P_j$ is a rank-one projection operator and $0\le\mu_j=\tr E_j\le1$.  The trace of the completeness relation implies that $\sum_j\mu_j={}$(dimension of the quantum system).  For a rank-one POVM, we have
\begin{equation}
	\label{eq:povm2}
	H(p_j)\geq S(\rho)-\sum_j p_j\log\mu_j\ge S(\rho)\;.
\end{equation}
The ensemble inequality~\cite{nielsen00a} says that the Shannon information of a set of ensemble probabilities $q_j$ exceeds the Holevo quantity of the ensemble:
\begin{equation}
H(q_j)\ge S\biggl(\sum_j q_j \rho_j\biggr)-\sum_j q_j S(\rho_j)\;.
\end{equation}

For strategy~(a), where  the local measurements are in the eigenbases of the marginal density operators, we have immediately that $H(A)=S(A)$ and $H(B)=S(B)$.  For both~(b) and (c), we can apply the POVM inequality in its rank-one form to $p_a=\tr(E_a\rho_A)$ to conclude that $H(A)\ge S(A)$.  Similarly, for strategy~(b), the POVM inequality applied to $p_b=\tr(F_b\rho_B)$ gives $H(B)\ge S(B)$.  For strategy~(c), we need a chain of inequalities to conclude that $H(B)\ge S(B)$:
\begin{align}
H(B)=H(C,B)&=H(C)+H(B|C)\nonumber\\
&=H(p_c)+\sum_c p_c H(B|c)\nonumber\\
\label{eq:first}
&\ge H(p_c)+\sum_c p_c S(\rho_{B|c})\\
&\ge S\biggl(\sum_c p_c\rho_{B|c}\biggr)
\label{eq:second}\\
&=S(\rho_B)=S(B)\;.
\end{align}
The first inequality~(\ref{eq:first}) is a consequence of applying the POVM inequality to Eq.~(\ref{eq:pbgivenc}), the second inequality~(\ref{eq:second}) is an example of the ensemble inequality, and the final equality uses Eq.~(\ref{eq:rhoBensembles}).

\subsubsection{Type~1: Mutual-information-based measures}

For type-1 measures, we choose $\cQ_1=S(A:B)$ and $\cC_1=H(A:B)$, giving
the difference measure
\begin{equation}
	\label{eq:type1}
	{\cal M}_{1} = S(A:B) - H(A:B)\;.
\end{equation}
We now apply the three measurement strategies introduced in Sec.~\ref{sec:intro} to obtain the classical mutual information $H(A:B)$; this leads to three different type-1 measures.

For strategy~(a), the local measurements are made in the eigenbases of the marginal density operators, and this gives a nonclassical-correlation measure that we denote by ${\cal M}_{1a}$.  If the marginal density operators have nondegenerate eigenvalues, the marginal eigenbases are unique; in the case of degeneracy, one needs to maximize $H(A:B)$ over the rank-one, projection-valued measurements in the degenerate subspaces to get a unique measure~$\cM_{1a}$.  The measure $\cM_{1a}$ was introduced by Luo in~\cite{luo08a} and called there the \textit{measurement-induced disturbance\/} (MID).  The same measure, in a different guise, had been proposed by Rajagopal and Randall in~\cite{rajagopal02a}; they defined what they called the \textit{quantum deficit\/} as $H(A,B)-S(A,B)$, where $H(A,B)$ is obtained from measurements in the marginal eigenbases.  The quantum deficit and MID are the same because they differ by the terms $H(A)-S(A)$ and $H(B)-S(B)$, which are zero for measurements in the marginal eigenbases.

When strategy~(b) is used, we obtain the measure
\begin{equation}
\cM_{1b}=S(A:B)-\max_{{\rm(b)}}H(A:B)\;,
 \end{equation}
where the classical mutual information has to be maximized over the unconditioned local measurements of strategy~(b).  The maximum classical mutual information was introduced in~\cite{Piani2008a} as a measure of classical correlations, and the same paper suggested $\cM_{1b}$ as a measure of nonclassical correlations.  This measure was investigated in detail by Wu, Poulsen, and M{\o}lmer (WPM) in~\cite{wu09a}, and we refer to it as the \textit{WPM measure\/} in this paper, while denoting it as $\cM_{1b}$.   The optimal unconditioned local measurements are not necessarily orthogonal-projection-valued.  An example of a case in which the maximization requires POVMs and not just projective measurements was given in~\cite{wu09a}; we review and extend this example in Appendix~\ref{app:projvsPOVM}.  In addition, the optimal local measurements do not generally occur in the marginal eigenbases, which implies that $\cM_{1a}^{{\rm(MID)}}\ge\cM_{1b}^{{\rm(WPM)}}$.

For strategy~(c), the classical mutual information $H(A:B)$ can be made arbitrarily large, thus allowing $\cM_{1c}$ to be arbitrarily negative.  This is easy to see by considering the extreme case of~(c) in which every outcome~$a$ leads to a different measurement on system~$B$; then, as noted in Sec.~\ref{sec:localmeasurements}, $H(A:B)=H(A)$, which can be as big as desired by giving the measurement on $A$ an arbitrarily large number of outcomes.  We conclude that $\cM_{1c}$ has nothing to do with quantifying nonclassical correlations, so we drop $\cM_{1c}$ from our array of possible measures.

\subsubsection{Type~2: Conditional-entropy-based measures}

For type-2 measures, we choose $\cQ_2=-S(B|A)$ and $\cC_2=-H(B|A)$.  The result is the difference measure
\begin{equation}
\cM_{2} = H(B|A) - S(B|A)\;.
\end{equation}
We notice immediately that
\begin{equation}
\cM_{2} = \cM_{1} + \big[ H(B) - S(B)\big] \geq \cM_{1}\;.
\end{equation}
This shows that a type-1 measure is always less than or equal to the type-2 measure that uses the same measurement strategy, with equality only when $B$ is measured in the marginal eigenbasis.

Measurements in the eigenbases of the marginal density operators have $H(B)=S(B)$, so for strategy~(a), we have $\cM_{2a}=\cM_{1a}$, and our measure is again MID.

Strategy~(b) gives the measure
\begin{equation}
\cM_{2b}=\min_{{\rm (b)}}H(B|A)-S(B|A)\;,
\end{equation}
where we have to minimize $H(B|A)$ over all unconditioned local measurements.  We can conclude from general considerations that $\cM_{1a}^{{\rm(MID)}}=\cM_{2a}^{{\rm(MID)}}\ge\cM_{2b}\ge\cM_{1b}^{{\rm(WPM)}}$.  Notice also that the unconditioned local measurements that minimize $H(B|A)$ need not be the same as those that minimize $H(A|B)$. This means that $\cM_{2b}$ is intrinsically asymmetric between subsystems $A$ and $B$ even though the measurement strategy is symmetric.

Strategy~(c) gives the measure
\begin{equation}
\cM_{2c}=\min_{{\rm (c)}}H(B|A)-S(B|A)\;.
\end{equation}
The POVM inequality immediately gives a bound on $H(B|A)$,
\begin{equation}\label{eq:HBgivenAbound}
H(B|A)=\sum_a p_a H(B|a)\ge\sum_ap_aS(B|a)\equiv H_{\{E_a\}}(B|A)\;.
\end{equation}
When we are allowed to make conditional measurements on $B$, the bound can be achieved by measuring $B$, for outcome~$a$, in the eigenbasis of $\rho_{B|a}$.  Hence, with the conditional measurements on $B$ specified, the minimization of the classical conditional entropy, $H(B|A)$, is reduced to choosing a measurement on $A$ that minimizes the conditional entropy $H_{\{E_a\}}(B|A)$:
\begin{equation}\label{eq:tildeHBgivenA}
\min_{{\rm(c)}} H(B|A) = \min_{\{E_a\}} H_{\{E_a\}}(B|A) \equiv \widetilde{H}(B|A)\;.
\end{equation}
The quantity $\widetilde{H}(B|A)$ is a special sort of classical conditional entropy.  The resulting measure is the \textit{quantum discord\/}~\cite{zurek00a,ollivier01a}:
\begin{equation}\label{eq:quantumdiscord}
	\cM_{2c} = \widetilde{H}(B|A) - S(B|A)\equiv{\sD}(A \rightarrow B)\;.
\end{equation}
Here we also introduce a notation for discord that emphasizes explicitly its asymmetry between $A$ and $B$.  In Appendix~\ref{app:projvsPOVM}, we exhibit joint states that show that to find the minimum $\widetilde H(B|A)$---and, hence, to find the quantum discord---sometimes requires rank-one POVMs, not just orthogonal-projection-valued measurements.

Henderson and Vedral~\cite{henderson01a} introduced the quantity $J(A\rightarrow B) = S(B) - \widetilde{H}(B|A)$ as a measure of classical correlations.  Ollivier and Zurek~\cite{ollivier01a} considered $J(A\rightarrow B)$ to be an asymmetric, measurement-based version of the mutual information and thus defined quantum discord as ${\sD}(A \rightarrow B) = S(A:B) - J(A\rightarrow B) = -S(B|A)+ \widetilde{H}(B|A)=\cM_{2c}$.  In particular, Ollivier and Zurek did not define discord in terms of conditioned measurements on $B$, but rather assumed that the quantity to be minimized over measurements on $A$ is the conditional entropy $H_{\{E_a\}}(B|A)$.

We can conclude from general considerations that $\cM_{1a}^{{\rm(MID)}}=\cM_{2a}^{{\rm(MID)}}\ge\cM_{2b}\ge\cM_{2c}^{{\rm(discord)}}$.  Our present considerations do not, however, provide an ordering of the WPM measure and quantum discord.  We return to the ordering of WPM and discord in Sec.~\ref{sec:summary} and show in Appendix~\ref{app:WPMdiscord} that $\cM_{1b}^{{\rm(WPM)}}\ge\cM_{2c}^{{\rm(discord)}}$.

\subsubsection{Type~3: Demon-based measures}

Type-3 measures quantify the difference in the work that can be extracted from a quantum system by quantum and classical demons.  The demons extract work by transforming the initial joint state $\rho_{AB}$ to the fully mixed joint state using any means at their disposal, including measurements.  We assume here that all states of the system have the same energy so that all the work that the demons extract arises from the entropy difference between the initial and final states of the system; it is natural to choose $k_{B}T \ln 2$ as the unit of work.  Throughout the paper, whenever we talk about extractable work and erasure cost, we actually mean average work and average erasure cost.

The maximum work that can be extracted by a quantum demon by any means is given by the entropy difference between the initial and final states,
\begin{equation}
	\label{eq:workQ}
	W_{q} = \log(d_{A} d_{B}) -S(A,B)\;,
\end{equation}
where $d_{A}$ and $d_{B}$ are the dimensions of the two subsystems.  The demon could extract this amount of work by devising an optimal process that directly transforms the joint state $\rho_{AB}$ to the maximally mixed state.  It could, instead, make a measurement in the joint eigenbasis of $\rho_{AB}$, extract work $\log(d_Ad_B)$ as the post-measurement pure eigenstate is transformed to the maximally mixed state, and then pay a price $S(A,B)$ to erase its memory of the $S(A,B)$ bits acquired in the measurement.  The demon would then be ready to pick up another copy of the system and repeat the process.

In contrast to a quantum demon, a local, classical demon can only manipulate the subsystem in its possession.  In Sec.~\ref{sec:intro} we introduced two cases for the local demons that are dealing with our bipartite system.  In case~(i) the two demons are not allowed to communicate with each other.  In this case, the maximum amount of work demon~$A$ can extract from subsystem~$A$ is $\log d_A - S(A)$.  This can be achieved by an optimal process that directly transforms the marginal state $\rho_A$ to the maximally mixed state or by measuring in the marginal eigenbasis, extracting work $\log d_A$ as the post-measurement pure state is transformed to the maximally mixed state, and then erasing the $S(A)$ bits of measurement record at cost $S(A)$.  Since demon~$B$ is in the same situation, the maximum work the two local demons can extract is
\begin{equation}
\label{eq:Wci}
W_{c} = \log(d_{A}d_{B}) - S(A) - S(B)\;.
\end{equation}
The difference in the amount of work that can be extracted by the quantum and classical demons, called the \textit{work deficit\/}~\cite{Oppenheim2002a,zurek03b,brodutch10a}, is the quantum mutual information:
\begin{equation}
	\label{eq:workdeficiti}
	W_{q} - W_{c}= S(A)+S(B)-S(A,B) = S(A:B)
    \equiv\cM_{{\rm 3(i)}}\;.
\end{equation}
Brodutch and Terno~\cite{brodutch10a} have noted that the work deficit in the case of erasure without communication between the local demons provides an operational interpretation of the quantum mutual information.

In case~(ii) the local demons can communicate their measurement results and thus reduce their cost of erasure.  In particular, the demons make local measurements, which in accord with the assumptions of this section are described by rank-one POVMs and thus leave the two subsystems in pure states.  They can then extract work
\begin{equation}
W^{+} = \log d_{A} + \log d_{B}
\end{equation}
as their respective systems are transformed to the maximally mixed state.  They must then erase their memories of the measurement record so they are ready to handle another copy of the joint state $\rho_{AB}$.

In the absence of communication, the total erasure cost is $W^{-}=H(A)+H(B)\ge S(A)+S(B)$, with the minimum attained for measurements in the marginal eigenbases; the net work the demons can extract is that of case~(i), i.e.,  $W_c=W^+ - W^-=\log{d_Ad_b}-S(A)-S(B)$.  If the demons can communicate, however, as in case~(ii), then they can take advantage of correlations between their measurement results to reduce their erasure cost to the joint classical information in their measurement records, $W^{-} = H(A,B)$, which gives net work
\begin{equation}\label{eq:Wc}
W_c=W^+ - W^-=\log(d_Ad_B)-H(A,B)\;.
\end{equation}
Thus in case~(ii), the work deficit becomes
\begin{equation}\label{eq:workdeficitii}
W_q-W_c=H(A,B)-S(A,B)=\cM_3\;,
\end{equation}
giving us joint-entropy-based measures of nonclassical correlations, with $\cQ_3=-S(A,B)$, $\cC_3=-H(A,B)$, and $\cM_3=\cQ-\cC$.

We now have to consider the three measurement strategies for the local demons, but before embarking on that, we note that
\begin{equation}
	\label{eq:m3b}
	\cM_{3} =  \cM_{2} + \big[H(A) - S(A) \big]
	= \cM_{1} + \big[H(B) - S(B) \big] + \big[ H(A) - S(A) \big]\;,
\end{equation}
so for each measurement strategy, we have $\cM_{3}\ge\cM_2\ge\cM_1$, as we have noted earlier.

For strategy~(a), measurement in the marginal eigenbases, we have $H(A)=S(A)$ and $H(B)=S(B)$, so we again get the MID measure, i.e., $\cM_{3a} = \cM_{2a} = \cM_{1a}$; this is the form in which Rajagopal and Randall~\cite{rajagopal02a} defined what they called the quantum deficit.  For strategy~(b), we have to minimize $H(A,B)$ over all unconditioned local measurements,
\begin{equation}\label{eq:M3b}
\cM_{3b}=\min_{{\rm (b)}}H(A,B)-S(A,B)\;;
\end{equation}
in general, the result is not the same as $\cM_{2b}$ or $\cM_{1b}$.

For strategy~(c), we have to minimize $H(A,B)$ over all conditioned local measurements.  The minimization over the conditioned measurements on $B$ is simple, since as in Eq.~(\ref{eq:HBgivenAbound}), we have
\begin{equation}\label{eq:HEAB}
H(A,B) = H(A) + H(B|A)\geq H(A) + \sum_a p_a S(B|a)\equiv H_{\{E_a\}}(A,B)\;,
\end{equation}
with equality if and only if the measurement on $B$, given outcome~$a$, is in the marginal eigenbasis of $\rho_{B|a}$.  Hence, with the conditional measurements on $B$ specified, the minimization of the classical joint entropy, $H(A,B)$, is reduced to choosing a measurement on $A$ that minimizes the joint entropy $H_{\{E_a\}}(A,B)$:
\begin{equation}\label{eq:tildeHAB}
\min_{{\rm(c)}} H(A,B) = \min_{\{E_a\}} H_{\{E_a\}}(A,B) \equiv \widetilde{H}(A,B)\;.
\end{equation}
The quantity $\widetilde{H}(A,B)$ is a special sort of classical joint entropy.  The resulting measure of nonclassical correlations is
\begin{equation}\label{eq:M3c}
\cM_{3c} = \widetilde{H}(A,B) - S(A,B)\;.
\end{equation}
This measure was hinted at in Zurek's original paper on discord~\cite{zurek00a}.  Ollivier and Zurek~\cite{ollivier01a} defined quantum discord as the quantity $\cM_{2c}$, but Zurek \cite{zurek03b} resurrected $\cM_{3c}$ as a modified form of discord in his paper on discord and Maxwell demons.  Brodutch and Terno~\cite{brodutch10a} have also pointed out that $\cM_{3c}$ is the measure that applies to demons that can communicate and use strategy~(c) for their measurements.  Hence, we can call $\cM_{3c}$ the \textit{demon discord\/}~(dd).

As noted in Sec.~\ref{sec:intro}, we have $\cM_{3a}^{{\rm MID}}\ge\cM_{3b}\ge\cM_{3c}^{{\rm(dd)}}$.

\subsubsection{Properties of nonclassical-correlation measures}
\label{sec:summary}

The following array neatly summarizes the measures of nonclassical correlations that we have found and the relations we have found among them:
\begin{equation}
	\label{eq:table}
	\begin{array}{ccccccc}
		& & \cM_{1a}^{({\rm  MID})} & \geq & \;\;\;\cM_{1b}^{({\rm WPM})} & & \\
		& & \rotatebox[origin=c]{90}{$=$}  & & \rotatebox[origin=c]{90}{$\geq$} &  & \\
		& & \cM_{2a}^{({\rm  MID})} & \geq & \cM_{2b} & \geq & \;\;\;\; \cM_{2c}^{({\rm  discord})} \\
		& & \rotatebox[origin=c]{90}{$=$}  & & \rotatebox[origin=c]{90}{$\geq$} &  & \rotatebox[origin=c]{90}{$\geq$} \\
		S(A,B)\;\; = \;\; \cM_{{\rm 3(i)}} & \geq \;\;& \cM_{3a}^{({\rm MID})} & \geq & \cM_{3b} & \geq & \cM_{3c}^{{\rm(dd)}}
	\end{array}
\end{equation}
The vertically oriented inequalities are best read by leaning your head to the left; in the absence of leaning, the wedges point toward the smaller quantity, as is standard.

Of the potential measures we started with, the demon-based measure that assumes erasure without communication is special and gives the quantum mutual information.  Of the remaining nine potential measures, we discarded one, $\cM_{1c}$, as meaningless; we found that the three measures in the left column of the array are all identical to the MID measure; we determined that three of the other measures are the WPM measure, quantum discord, and demon discord; and we are thus left with two new measures, $\cM_{2b}$ and $\cM_{3b}$, although $\cM_{3b}$ is very closely related to---and perhaps identical to---a discord-like measure introduced by Modi~{\it et al.}~\cite{modi10a}.

Modi~{\it et al.}~\cite{modi10a} introduced a set of measures of quantum and classical correlations based on the relative-entropy distance~(\ref{eq:qrelativeentropy}) between a multi-partite state $\rho$ and the nearest state $\sigma_\rho$ that is diagonal in a product basis, or between $\rho$ and the nearest product state.  The only one of these measures relevant to our discussion is their ``discord,'' which when specialized to bipartite states, is the distance $\sD_{\rm Modi}=\min_{\sigma_{AB}}S(\rho_{AB}||\sigma_{AB})$, where $\sigma_{AB}$ is diagonal in a product basis.  Modi {\it et al.}\ show that the minimum is attained on a state obtained by projecting $\rho_{AB}$ into a product basis, i.e., $\sigma_{AB}=\sum_{a,b}|e_a,f_b\rangle\langle e_a,f_b|\rho_{AB}|e_a,f_b\rangle\langle e_a,f_b|$, in which case, $S(\rho_{AB}||\sigma_{AB})=S(\sigma_{AB})-S(\rho_{AB})$.  Thus we have
\begin{equation}
\sD_{\rm Modi}=\min_{\{|e_a,f_b\rangle\}}S(\sigma_{AB})-S(\rho_{AB})\;.
\end{equation}
Since $S(\sigma_{AB})$ is the classical joint entropy of a measurement made on $\rho_{AB}$ in the product basis $|e_a,f_b\rangle$, this would be the same as our $\cM_{3b}$ if we knew that the optimal local measurements for $\cM_{3b}$ were described by \textit{orthogonal\/} rank-one projectors.

Brodutch and Terno~\cite{brodutch10a} define three kinds of ``discord'': their $D_1$ is the standard discord $\cM_{2c}^{{\rm(discord)}}$; their $D_2$ is the demon discord $\cM_{3c}^{{\rm(dd)}}$; and their $D_3$ is a discord-like quantity that uses a different conditional measurement strategy.  This strategy allows conditioned local measurements, but with the measurement on $A$ constrained to be in the marginal eigenbasis of $\rho_A$.  The Brodutch-Terno measurement strategy is a restriction of strategy~(c), and (a) is a restriction of the Brodutch-Terno strategy.  Measures based on it could thus be placed in the array~(\ref{eq:table}) as an alternative intermediate column whose ordering with strategy~(b) is indeterminate.

All the measures in the array~(\ref{eq:table}), except the quantum mutual information, are bounded above by MID, and MID is bounded above by the quantum mutual information.  Similarly, MID, $\cM_{2b}$, and $\cM_{3b}$ are bounded below by both the WPM measure and the quantum discord, and the demon discord $\cM_{3c}$ is bounded below by discord.  The WPM measure and quantum discord have a special status in that they are the most parsimonious of the measures in quantifying nonclassical correlations.

WPM showed that their measure is nonnegative, and Datta~\cite{datta08b} showed that discord is nonnegative, allowing us to conclude that all the other measures are also nonnegative.  Both proofs rely on the strong subadditivity of quantum entropy~\cite{nielsen00a}; we review the proofs in Appendix~\ref{app:WPMdiscord}.  Careful consideration of the conditions for saturating the strong-additivity inequality~\cite{Hayden2004a}, not presented here, give the conditions for WPM and discord to be zero: the WPM measure is zero if and only if $\rho_{AB}$ is diagonal in a \textit{product basis}, i.e., an orthonormal basis of the form $|e_a\rangle\otimes |f_b\rangle$, and discord is zero if and only if $\rho_{AB}$ is diagonal in a \textit{conditional product basis\/} (pointing from $A$ to $B$), i.e., an orthonormal basis of the form $|e_a\rangle\otimes|f_{b|a}\rangle$.

Since MID, like WPM, is zero if and only if $\rho_{AB}$ is diagonal in a product basis, the relations in the array~(\ref{eq:table}) imply that $\cM_{2b}$ and $\cM_{3b}$ are zero if and only if $\rho_{AB}$ is diagonal in a product basis.  Similarly, the inequality $\cM_{3c}^{{\rm(dd)}}\ge\cM_{2c}^{{\rm(discord)}}$ shows that having $\rho_{AB}$ diagonal in a conditional product basis is necessary to make $\cM_{3c}$ zero, and a moment's contemplation of Eqs.~(\ref{eq:HEAB})--(\ref{eq:M3c}) shows that this is also a sufficient condition.

For pure states, we have $S(A,B)=0$, $S(A)=S(B)=-S(B|A)=-S(A|B)$, and $S(A:B)=2S(A)=2S(B)$.  It is easy to show that the optimal measurement for all the measures in the array is measurement in the Schmidt basis of the pure state (marginal eigenbasis for each subsystem), which gives $H(A)=H(B)=H(A,B)=H(A:B)=S(A)=S(B)$ and $H(B|A)=H(A|B)=0$.  Thus all the measures in the array, except the quantum mutual information, are equal to the marginal quantum entropy, $S(A)=S(B)$, which is the entropic measure of entanglement for bipartite pure states.

The remaining gap in our understanding left by the relations in the array is whether there is an inequality between the WPM measure and discord.  The WPM measure is strictly bigger than zero for states that are diagonal in a conditional product basis that is not a product basis and so is bigger than the quantum discord for such states. If there is an inequality, it must be that the WPM measure is bounded below by quantum discord.  Indeed, it is not hard to come up with a proof, using the method of Piani~\textit{et al.}~\cite{Piani2008a}.  The proof, given in Appendix~\ref{app:WPMdiscord}, is part of the two-step demonstration that WPM and discord are nonnegative.  We conclude that
\begin{equation}
\cM_{1b}^{{\rm(WPM)}}\ge\cM_{2c}^{{\rm(discord)}}=\sD(A\rightarrow B)\;.
\end{equation}
The proof allows us to identify the equality condition: the WPM measure is equal to discord if and only if $\rho_{AB}$ is diagonal in a conditional product basis that points from $B$ to $A$.

We emphasize that product bases and conditional product bases do not exhaust the set of orthonormal bases that are made up of product states.  There are orthonormal bases made up entirely of product states that are neither product bases nor conditional product bases; these have been studied, for example, in the context of nonlocality without entanglement~\cite{Bennett1999a}.  Not surprisingly, we refer to such a basis as a basis of product states, to be distinguished from a product basis or a conditional product basis.

\subsection{Rank-one POVMs and projective measurements}
\label{sec:POVMs}

In Sec.~\ref{sec:Measures} we assumed that all the measurements were described by rank-one POVMs.  This assumption does not affect the demon-based work deficit~(\ref{eq:workdeficiti}) for the case of erasure without communication, for that case, which leads to the quantum mutual information, does not rely on any assumptions about how the subsystems are measured.  Nor does this assumption affect MID, which is derived from measurement strategy~(a), a strategy that from the outset prescribes orthogonal-projection-valued measurements in the eigenbases of the marginal density operators.  The assumption must be carefully examined, however, for measurement strategies~(b) and~(c).  On the face of it, there is a problem for the second and third rows of our array.  For type-2 measures, the task is to minimize a classical conditional entropy, and for type-3 measures, the task is to minimize a classical joint entropy.  In both cases, the minimum is achieved by making no measurements at~all.

For the demon-based measures in the right two columns of the third row, it is clear what the problem is.  The contribution of $H(A,B)$ to the classical work comes from the erasure cost; the local demons can minimize their erasure cost by not having a measurement record.  Of course, if the local demons make no measurements, they also cannot extract the work attendant on knowing more about their system's state.  The upshot is that formula~(\ref{eq:Wc}) for the net classical work needs to be modified if one does not assume measurements described by rank-one POVMs.  Appendix~\ref{app:demons} shows, not surprisingly, that, once modified, the net classical work is always optimized on rank-one POVMs, so one can restrict the demons in this way without affecting their performance.

For the measure $\cM_{2b}$, we know of no reason to restrict to rank-one POVMs more compelling than declaring that the measure would be nonsense without this restriction.  For the quantum discord, we can do better: the original definition of discord~\cite{ollivier01a} did not discuss conditioned measurements on $B$, but rather formulated the discord directly in terms of minimizing the classical conditional entropy as in Eqs.~(\ref{eq:HBgivenAbound})--(\ref{eq:quantumdiscord}); this is equivalent to our assuming rank-one POVMs for the measurement on $B$.
We are still left with a question---why should the measurements on $A$ be restricted to rank-one POVMs?---and this same question applies to both local measurements for the WPM measure.  We now address this question by showing in both situations that the optimum can always be attained on rank-one POVMs.  It is important to show this, because the proofs regarding nonnegativity and ordering of the WPM measure and discord, given in Appendix~\ref{app:WPMdiscord}, assume rank-one POVMs.

We deal with the WPM measure first.  The key point is obvious: making coarse-grained POVM measurements on $A$ and $B$ should not uncover as much mutual information as making fine-grained, rank-one POVM measurements.  We start with POVMs $\{E_a\}$ and~$\{F_b\}$ for systems~$A$ and~$B$, and we imagine that these are a coarse graining of POVMs $\{E_{aj}\}$ and $\{F_{bk}\}$, i.e.,
\begin{equation}
E_a=\sum_j E_{aj}\;,\qquad F_b=\sum_kF_{bk}\;.
\end{equation}
A POVM element can always be fine-grained to the rank-one level by writing it in terms of its eigendecomposition.  The joint probability for the fine-grained outcomes~$aj$ and~$bk$ is
$p_{ajbk}=p_{jk|ab}p_{ab}$, with similar relations for the marginals for the two subsystems.  It is now trivial to show that fine graining never decreases the classical mutual information:
\begin{equation}
H(A,J:B,K)=H(A:B)+\sum_{a,b}p_{ab}H(J:K|a,b)\;.
\end{equation}
This means that in maximizing the classical mutual information, we need only consider rank-one POVMs.

For the discord, the reduction to rank-one POVMs has been demonstrated by Datta~\cite{datta08b}; it is sufficiently brief that we repeat it here.  Since the conditional measurements on $B$ are already specified, we need only worry about fine graining the measurement on~$A$.  We need the conditional state of $B$ given the coarse-grained outcome $a$ in terms of the conditional states given the fine-grained outcome $aj$:
\begin{equation}
\rho_{B|a}=\frac{\tr_A(E_{a}\rho_{AB})}{p_a}
=\sum_j\frac{\tr_A(E_{aj}\rho_{AB})}{p_a}=\sum_jp_{j|a}\rho_{B|aj}\;,
\end{equation}
where
\begin{equation}
p_{j|a}=\frac{p_{aj}}{p_a}=\frac{\tr_A(E_{aj}\rho_A)}{\tr_A(E_a\rho_A)}\;.
\end{equation}
The quantity to be minimized over measurements on $A$ is the conditional entropy~(\ref{eq:HBgivenAbound}).  For it, we can write
\begin{align}
H_{\{E_a\}}(B|A)&=\sum_ap_aS(\rho_{B|a})\nonumber\\
&=\sum_a p_a S\Big(\sum_j p_{j|a}\rho_{B|aj}\Big)\\
&\geq\sum_{a,j} p_{aj}S(\rho_{B|aj})=H_{\{E_{aj}\}}(B|A)\;,
\end{align}
where the inequality follows from the concavity of the von Neumann entropy.  Thus fine graining never increases this conditional entropy, so we are assured that the minimum is attained on rank-one POVMs.

We have now settled the question of restricting to rank-one POVMs for all the measures except the measure in the middle, $\cM_{2b}$, and for it, we simply assert that it makes sense only if we restrict to rank-one POVMs.  A remaining question is whether we can further restrict to orthogonal-projection-valued measurements.  Searching over the entire set of rank-one POVMs is a daunting task, considerably more onerous than searching just over projection-valued measurements.  On this score, we can report that WPM drew attention to an example where the WPM measurements are optimized on a rank-one POVM that is not projection-valued; we extend this example to quantum discord and generalize it in Appendix~\ref{app:projvsPOVM}.

These examples, however, require that at least one system have dimension bigger than two; For evaluating quantum discord for a two-quibit system, Chen \textit{et al.} \cite{chen2011} found, that there are some states for which three-element POVMs on system $A$ do better than two-outcome, orthogonal-projection-valued measurements. Exploring the situation numerically, Galve, Giorgi, and Zambrini \cite{galve2011} confirmed this finding, but suggested that the corrections to the two-qubit discord obtained by using POVMs, instead of orthogonal projectors, are negligible.  In the next section, we do a wholesale evaluation of  the various measures for two-qubit states; in the need for manageable numerics, we restrict the search over measurements to rank-one, orthogonal projection operators; according to \cite{galve2011}, this should have no significant effect on the result.

Several groups of investigators have considered Gaussian versions of nonclassical-correlation measures for Gaussian states of two harmonic-oscillator modes; the local measurements are restricted to Gaussian measurements, i.e., measurements whose POVM elements are the phase-space displacements of a particular single-mode Gaussian state.   Giorda and Paris~\cite{giorda2010a} and Addesso and Datta~\cite{adesso2010a} focused on a Gaussian version of discord and showed that the optimal Gaussian measurements are rank-one POVMs, but that for some Gaussian states, the optimal measurement is not orthogonal-projection-valued and thus requires the use of POVMs.  {Mi{\u s}ta}~\textit{et al.}~\cite{mista2011a} investigated Gaussian versions of MID and WPM (which they called AmeriolatedMID); their investigation showed that for some Gaussian states, the optimal Gaussian measurement for the WPM measure is not the globally optimal measurement when one allows nonGaussian POVMs.

\section{Numerical results for two-qubit states}
\label{sec:numerics}

\begin{figure}[htbp]
\begin{center}
\includegraphics[width=9cm]{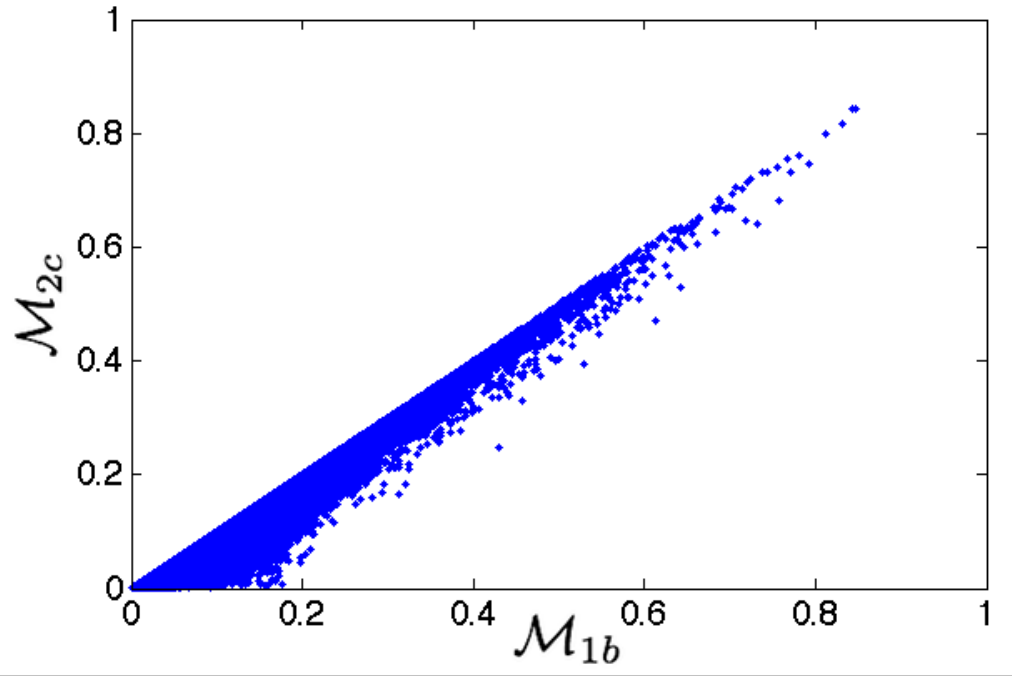}\\ 
\end{center}
\caption{$\cM_{2c}^{{\rm(discord)}}=\sD(A\rightarrow B)$ plotted against $\cM_{1b}^{{\rm(WPM)}}$ for one million randomly generated joint density matrices, using orthogonal projectors for the measurements.  As expected, the WPM measure is never smaller than the discord; also evident is that  discord is zero for a larger class of states than the WPM measure, those being the states that are diagonal in a conditional product basis pointing from~$A$ to $B$. \label{fig:relation1}}
\end{figure}

\begin{figure}[htbp]
\begin{center}
\includegraphics[width=9cm]{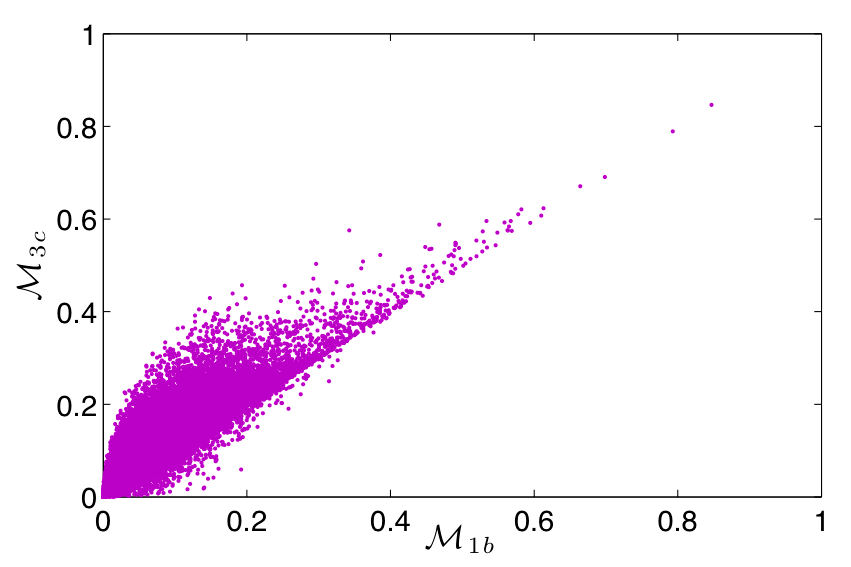}\\ 
\end{center}
\caption{$\cM_{3c}^{{\rm(dd)}} $ plotted against $\cM_{1b}^{{\rm(WPM)}}$ for 100,000 randomly generated joint density matrices. Since $\cM_{3c}^{{\rm(dd)}}\geq\cM_{2c}^{{\rm(discord)}}$, the points from Fig.~\ref{fig:relation1} move upwards.  Many points pass the diagonal, and the ordering of Fig.~\protect\ref{fig:relation1} disappears.\label{fig:relation2}}
\end{figure}

\begin{figure}[htbp]
\begin{center}
\includegraphics[width=9cm]{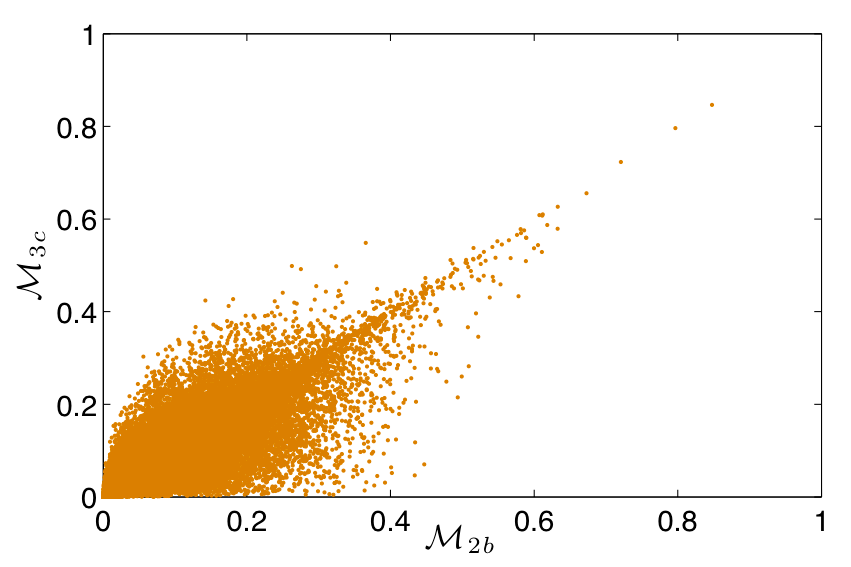}\\ 
\end{center}
\caption{$\cM_{3c}^{{\rm(dd)}}$ plotted against $\cM_{2b}$ for 100,000 randomly generated joint density matrices.  Relative to Fig.~\ref{fig:relation2}, the points move right, due to the relation $\cM_{2b} \geq \cM_{1b}$.  Since not all of them pass the diagonal, there is no ordering relation between  $\cM_{3c}$ and $\cM_{2b}$. \label{fig:relation3}}
\end{figure}

One purpose  of our framework is to clarify relations among the various measures of nonclassical correlations beyond entanglement.  The ordering of the measures is of particular interest.  The framework provides by construction some ordering relations between the measures; in addition, we have proved, using the method of Piani~\textit{et al.}, the important relation that the WPM measure is bounded below by the discord.  Nonetheless, questions remain, in particular, of whether there is an ordering between $\cM_{2b}$ and $\cM_{3c}^{{\rm(dd)}}$, as well as between $\cM_{1b}^{{\rm(WPM)}}$ and $\cM_{3c}^{{\rm(dd)}}$.

In this section we illustrate and investigate the various orderings by presenting numerical evaluations of the several measures for randomly selected two-qubit states.   It should be noted, however, that in order to do the optimizations over measurements numerically, we have had to restrict ourselves to orthogonal projectors instead of the more general POVMs, so in some situations, we might not be finding the optimal measurements.


To calculate the various correlation measures, we follow the approach of Al-Qasimi and James~\cite{qasimi}.  The measurement operators $E_a$ and $F_b$ are orthogonal projectors,
\begin{align}
E_a&=|e_a^A\rangle\langle e_a^A|\;,\qquad F_b=|e_b^B\rangle\langle e_b^B|\;,\qquad a,b \in \{0,1\}\;, \\
&|e_0^X\rangle=\cos\theta^X|0\rangle+e^{i\phi^X}\sin\theta^X |1\rangle\;,\quad
|e_1^X\rangle=-\sin\theta^X|0\rangle+e^{i\phi^X}\cos\theta^X|1\rangle\;.
\end{align}
The required optimization is done by a numerical search over the angles $\{\theta^X,\phi^X\}$ for $X \in \{A,B\}$.  For measurement strategy~(b), we must search over the four angles for both qubits, but for strategy~(c), we need only search over the two angles for subsystem~$A$.

Figure~\ref{fig:relation1} compares the WPM measure and discord, confirming the expectation that the WPM measure is never smaller than discord.  Figures~\ref{fig:relation2} and~\ref{fig:relation3} display the aforementioned pairs of correlation measures where our framework does not imply an ordering relation; the numerical data show that there is no ordering for these pairs.

\begin{figure}[htbp]
\begin{center}
\includegraphics[width=16cm]{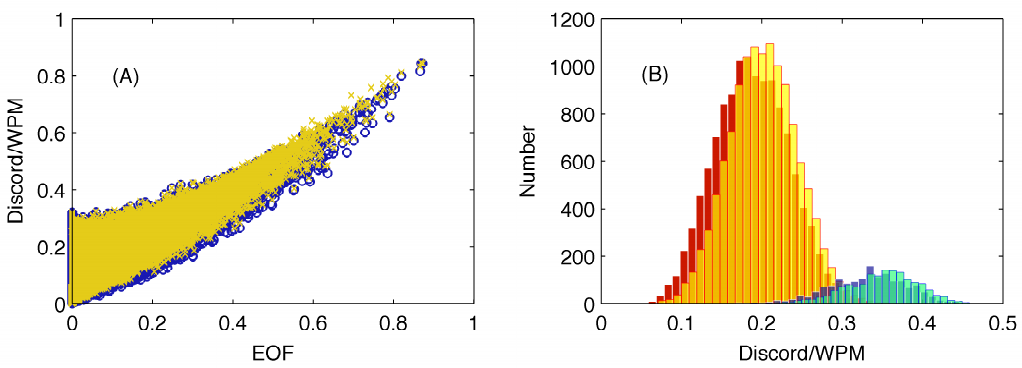}\\ 
\end{center}
\caption{(A) Discord (blue circles) and the WPM measure (yellow crosses) for one million randomly chosen two-qubit states, plotted against entanglement of formation, $E_f$.  As the correlations increase, the spread between entanglement and WPM or discord decreases. (B) Two superimposed histograms showing the distribution of discord and the WPM measure for ranges of values of $E_{f}$: left histogram shows discord (red) and WPM (yellow) for the states of (A) corresponding to  $0.1 \leq E_{f} \leq 0.2$; right histogram shows discord (blue) and WPM (green) corresponding to $0.3 \leq E_{f} \leq 0.4$.   \label{fig:eofrelation}}
\end{figure}

\begin{figure}[htbp]
\begin{center}
\includegraphics[width=8cm]{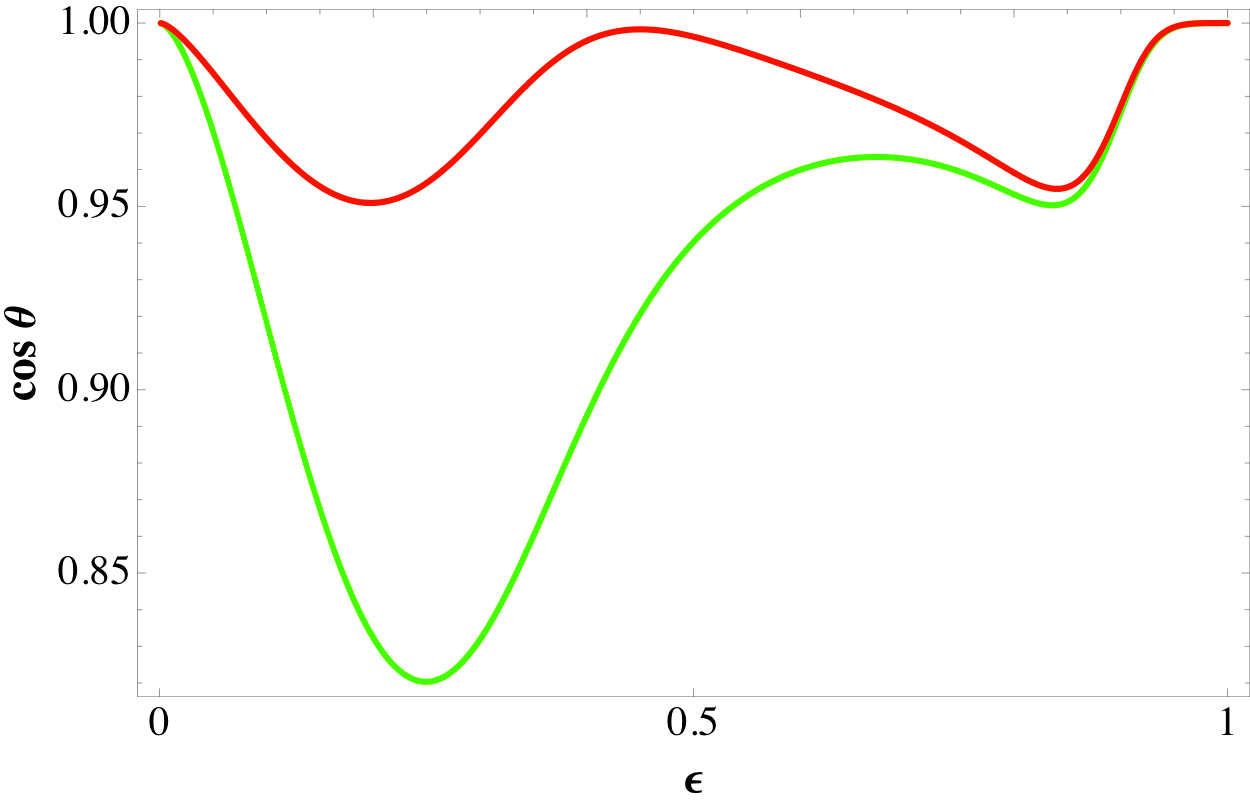}\\ 
\end{center}
\caption{Deviation of the numerically obtained, optimal measurement vectors from the maximal singular vectors of the correlation matrix for the WPM measure.  The joint states are a mixture of a pure product state with marginal spin (Bloch) vectors $\bm{a}=(1,\, 0,\, 0)$ and $\bm{b} = (1/\sqrt{2},\,-1/2,\,1/2)$ and a mixed Bell-diagonal (zero marginal spin vectors) state, with correlation matrix $c={\rm diag}(-0.9,\,-0.8,\,-0.7)$.  The mixing parameter is $\epsilon$, with $\epsilon=0$ corresponding to the product state and $\epsilon=1$ to the Bell-diagonal state.  The green curve shows the cosine of the angle between the maximal right singular vector and the measurement vector on system~$B$. The red curve is the cosine of the angle between maximal left singular vector and the measurement vector on system~$A$.   \label{fig:angles}}
\end{figure}

Another relation we have explored numerically is the one between the correlation measures and entanglement.  Figure~\ref{fig:eofrelation}  shows discord and the WPM measure plotted against entanglement of formation, reproducing the plot in~\cite{qasimi} for discord, but providing new data for the WPM measure.  The entanglement of formation is calculated using Wootters's analytical expression~\cite{wooters_eof}, $E_f(\rho)=h\bigl((1+\sqrt{1-\mathcal{C}^2(\rho)})/2\bigr)$.
Here $h(x)$ is the binary entropy, $h(x)=-x\log x-(1-x) \log(1-x)$, and $C(\rho)$ is the concurrence, given by $\mathcal{C}(\rho)=\max(0,\lambda_1-\lambda_2-\lambda_3-\lambda_4) $, where the $\lambda_j$s are the eigenvalues in decreasing order of the operator $\sqrt{\sqrt{\rho}\tilde{\rho}\sqrt{\rho}}$, with $\tilde{\rho}= (\sigma_y \otimes \sigma_y) \rho^* (\sigma_y \otimes \sigma_y)$.

To avoid the slow numerical optimization procedures, analytical expressions for the correlation measures would be desirable.  Yet only for very restricted classes of joint states are such expressions available. Girolami \textit{et al.}~\cite{amid} \footnote{In the published version of their paper  Girolami \textit{et al.} restricted the validity of their analytical expression to the set of two-qubit X-states \cite{girolami2011}.} suggested that there is an analytical expression for the WPM measure for general two-qubit states.  To understand their claim, we write the joint two-qubit state in terms of Pauli operators:
\begin{align}
\rho_{AB} =
\frac{1}{4}\Bigl(I_{AB} + \bm{a\cdot\sigma}^A\otimes I_B
+ I_A\otimes\bm{b\cdot\sigma}^B+
\sum_{j,k=1}^3 c_{jk}\sigma_j^A\otimes\sigma_k^B\Bigr)\;.
\end{align}
The correlation matrix, $c_{jk}=\tr(\sigma_j^A\otimes\sigma_k^B\rho_{AB})$, is not symmetric, but $c^Tc$ can be diagonalized as $c^Tc\bm{m_j}=\lambda_j^2\bm{m}_j$.  The eigenvalues, $\lambda_j^2$, are the squares of the singular values of $c$, and the eigenvectors are the right singular vectors of $c$.  The correlation matrix maps the right singular vectors to the left singular vectors, $c\bm{m}_j=\lambda_j\bm{n}_j$, and the left singular vectors, $\bm{n}_j$, are the eigenvectors of $cc^T$.

The claim of Girolami~\textit{et al.} was that the maximal left and right singular vectors, i.e., those corresponding to the largest singular value of $c$, specify the optimal measurements for the WPM measure.  We can confirm that the measurement vectors for generic (randomly generated) two-qubit states are close to the maximal singular vectors of $c$ for all three correlation measures that are based on measurement strategy~(b), but it can be shown analytically that in general the singular vectors are not the optimal measurement vectors.  Moreover, there are examples where the deviation becomes obvious in the numerics.  Figure~\ref{fig:angles} shows an example where the angle between the measurement vectors and the maximal singular vectors is noticeable in the calculation of the WPM measure.  Similar plots can be obtained for the measures $\cM_{2b}$ and $\cM_{3b}$.

\section{Conclusion}
\label{sec:conclusion}

We consider in this paper several entropic measures of nonclassical correlations.  All of these measures purport to quantify the degree of nonclassicality in bipartite quantum states.  Just as important as the degree of nonclassicality, however, is the boundary these measures set between quantum and classical states.  For the measures $\cM_{1a}^{{\rm(MID)}}$, $\cM_{1b}^{{\rm(WPM)}}$, $\cM_{2b}$, and $\cM_{3b}$, a joint state is classical if and only if it is diagonal in a product basis; this notion of classicality is clearly a consequence of using measurement strategies~(a) and~(b).  The measures $\cM_{2c}^{{\rm(discord)}}$ and $\cM_{3c}^{{\rm(dd)}}$ both consider a joint state to be classical if it is diagonal in a conditional product basis pointing from $A$ to $B$; this boundary between quantum and classical is a consequence of measurement strategy~(c).  The measures differ in the ``amount'' of nonclassicality they assign to states they deem not classical.

The boundary between quantum and classical is important by itself, first, because it can usually be extended to multipartite systems even when the measure of nonclassical correlations is not so easily extended and, second, because it serves as the basis for interesting questions about quantum-information processing.  For example, Eastin~\cite{Eastin2010au} has recently investigated whether \textit{concordant\/} computations can be simulated classically.  A concordant computation is one such that after every elementary gate, the state of the whole computer is diagonal in a product basis.  The entire computation is just a matter of changing the product basis, yet Eastin finds it difficult in general to simulate such computations efficiently.

This discussion raises at least two other questions, the first of which concerns the boundary between quantum and classical.  The boundaries induced by the nonclassical-correlation measures considered here are natural---na{\"\i}ve is perhaps a better word---in that the classicality of a state is defined in terms of properties of its eigenvectors.  This is quite different from the distinction between separable and entangled states~\cite{Werner1989a}, which pays no attention to the properties of the joint state's eigenvectors; a state is separable if and only if it has an ensemble decomposition---not an eigendecomposition---in terms of product states.  The boundaries for the measures discussed in this paper do require a classical state to have unentangled eigenvectors, but they are more restrictive than saying that a state is classical if its eigenvectors are product states.  There are orthonormal bases of product states that are neither product bases nor conditional product bases~\cite{Bennett1999a}.  The boundaries considered here, imposed by the measurement strategies, clearly have to do with wanting the product states in a joint eigenbasis to persist into the marginal eigenbasis of one or both subsystems.  A natural question is wheher there is some other way of setting the quantum-classical boundary so that a joint state is classical if and only if it has unentangled eigenvectors?

Any measure of nonclassical correlations assigns a number to a joint quantum state; the second question has to do with what this ``amount'' means. The demon-based measures have an operational interpretation as the work deficit suffered by local demons that are required to work only with the subsystems.  Recently, there have been two closely related proposals for operational interpretations of quantum discord in terms of state-merging protocols~\cite{DCavalcanti2011a,Madhok2011a}. Quantum discord has also been connected to the entanglement loss when mixed states are created from entangled states followed by entanglement distillation from those mixed states~\cite{cornelio2010a}.  Such operational interpretations are essential to understand the meaning of a measure of nonclassical correlations.  Whenever a measure of nonclassical correlations is proposed, the amount of nonclassicality must ultimately gain meaning through an operational interpretation.

\acknowledgments
We thank M.~Piani for pointing out to us how to prove that discord is never larger than the WPM measure~\cite{Piani2008a} and S. Vinjanampathy for apprising us of \cite{chen2011}. This work was supported in part by US National Science Foundation Grant Nos.~PHY-1005540 and PHY-0903953.  AS acknowledges the support of the Department of Science and Technology, Government of India, through the Ramanujan Fellowship (SR/S2/RJN-01/2009). AS also acknowledges the hospitality provided by the Center for Engineered Quantum Systems, University of Queensland, during the summer of 2010, where part of this work was done.

\appendix

\section{The POVM inequality}
\label{app:POVMinequality}

A quantum state written in its eigenbasis,
\begin{equation}
\rho = \sum_{\alpha} \lambda_{\alpha} |e_{\alpha} \rangle \langle e_{\alpha}|
\;,
\end{equation}
is subjected to a POVM with elements $E_{j}$.  This gives outcome probabilities
\begin{equation}
p_{j} = {\rm tr}(E_j\rho) =
\sum_{\alpha} \lambda_{\alpha} q_{j|\alpha}\;,
\end{equation}
where $q_{j|\alpha}=\langle e_\alpha|E_j|e_\alpha\rangle$ is the probability for outcome~$j$ given state $|e_\alpha\rangle\langle e_\alpha|$.  We define
\begin{equation}
\mu_j\equiv\tr E_{j}=\sum_{\alpha} q_{j|\alpha}\;,
\end{equation}
which implies that $q_{j|\alpha}/\mu_j$ is a normalized probability distribution on $\alpha$.

Now let $f(x) = - x \log x$ and proceed through the following steps:
\begin{align}
	H(p_j) & = \sum_{j} f(p_{j})\nonumber \\
	& = \sum_{j} f \bigg( \sum_\alpha \frac{q_{j|\alpha}}{\mu_j} \lambda_{\alpha}\mu_j\bigg) \nonumber \\
	& \geq \sum_{\alpha, j} \frac{q_{j|\alpha}}{\mu_j} f ( \lambda_{\alpha}\mu_j ) \label{eq:a1}\\
	& = - \sum_{\alpha, j} q_{j|\alpha}\lambda_{\alpha}
\log( \lambda_{\alpha}\mu_j) \nonumber \\
	& = - \sum_{\alpha, j} q_{j|\alpha} \lambda_\alpha \log\lambda_{\alpha}
- \sum_{\alpha, j} \lambda_{\alpha} q_{j|\alpha} \log \mu_j \nonumber \\
	& = - \sum_{\alpha} \lambda_{\alpha} \log \lambda_{\alpha} - \sum_{j} p_{j} \log \mu_j \label{eq:a2} \\
	& = S(\rho) - \sum_{j} p_{j}\log (\tr E_{j})\;.
\label{eq:a3}
\end{align}
The inequality (\ref{eq:a1}) uses that $f(x)$ is a concave function and that $q_{j|\alpha}/\mu_j$ is a probability distribution over $\alpha$.  The step leading to Eq.~(\ref{eq:a2}) uses that  $q_{j|\alpha}$ is a normalized distribution over~$j$.

\section{Nonnegativity and ordering of the WPM measure and discord }
\label{app:WPMdiscord}

In this Appendix we show that the WPM measure is bounded below by quantum discord and that the quantum discord is nonnegative.  We proceed by assuming that systems~$A$ and~$B$ are subjected to measurements described by POVMs with rank-one POVM elements $E_a$ and $F_b$, as in measurement strategy~(b).  For convenience, we use the fact that any set of POVM elements can be extended to rank-one orthogonal projection operators in a space of higher dimension, an extension called the Naimark extension~\cite{naimark40a}.  Formally, we have an orthonormal basis $|e_a\rangle$ in the higher-dimensional space such that $E_a=P_A|e_a\rangle\langle e_a|P_A$, where $P_A$ projects onto the original Hilbert space of system~$A$, which is the space where the states of system~$A$ live; similarly, there is an orthonormal basis $|f_b\rangle$ such that $F_b=P_B|f_b\rangle\langle f_b|P_B$, where $P_B$ projects onto the original Hilbert space of system~$B$.

We can write the joint probability for outcomes~$a$ and~$b$ as
\begin{equation}
p_{ab}=\tr(E_a\otimes F_b\rho_{AB})=
\tr\Bigl(P_A|e_a\rangle\langle e_a|P_A
\otimes P_B|f_b\rangle\langle f_b|P_B\rho_{AB}\Bigr)=
\langle e_a,f_b|\rho_{AB}|e_a,f_b\rangle\;,
\end{equation}
where the last equality follows because $\rho_{AB}$ lives in the original Hilbert space of $A$ and $B$, so we can discard the projectors into that space.  Other results we need below include
\begin{equation}
\rho_{B|a}=\frac{\tr_A(E_a\rho_{AB})}{p_a} = \frac{\langle e_a|\rho_{AB}|e_a\rangle}{p_a}\;,
\end{equation}
where
\begin{equation}
p_a=\tr(E_a\rho_A)=\langle e_a|\rho_A|e_a\rangle\;.
\end{equation}

We now extend the joint state~$\rho_{AB}$ to a space with two additional systems, $C$ and $D$.  We let $C$ have dimension equal to the number of outcomes~$a$, with an orthonormal basis $|g_a\rangle$, we let $D$ have dimension equal to the number of outcomes~$b$, with orthonormal basis $|h_b\rangle$.  The extended state,
\begin{equation}
\rho'_{ABCD}=\sum_{a,a',b,b'}
|e_a,f_b\rangle\langle e_a,f_b|
\rho_{AB}
|e_{a'},f_{b'}\rangle\langle e_{a'},f_{b'}|
\otimes|g_a\rangle\langle g_{a'}|\otimes|h_b\rangle\langle h_{b'}|\;,
\end{equation}
can be regarded as a state where systems~$C$ and~$D$ record the measurement outcomes in their orthonormal bases.  Notice that the entropy of the extended state is
\begin{equation}
S'(A,B,C,D)=S(A,B)\;.
\end{equation}

The proof follows from two applications of the strong-subadditivity property of von Neumann entropy~\cite{nielsen00a}.  Various marginals of the extended state and their von Neumann entropies enter into the proof:
\begin{align}
&\rho'_{ABD}=\sum_{a,b,b'}p_a|e_a\rangle\langle e_a|
\otimes
|f_b\rangle\langle f_b|\rho_{B|a}|f_{b'}\rangle\langle f_{b'}|
\otimes|h_b\rangle\langle h_{b'}|\;,\\
&\rho'_{AC}=\sum_{a,a'}
|e_a\rangle\langle e_a|
\rho_A
|e_{a'}\rangle\langle e_{a'}|
\otimes|g_a\rangle\langle g_{a'}|\;,\qquad
\rho'_{BD}=\sum_{b,b'}
|f_b\rangle\langle f_b|
\rho_B
|f_{b'}\rangle\langle f_{b'}|
\otimes|h_b\rangle\langle h_{b'}|\;,\\
&\rho'_{AB}=\sum_{a,b}p_{ab}|e_a,f_b\rangle\langle e_a,f_b|\;,\qquad
\rho'_A=\sum_{a}p_a|e_a\rangle\langle e_a|\;,\qquad
\rho'_B=\sum_{a}p_b|f_b\rangle\langle f_b|\;,\\
&\rho'_C=\sum_{a}p_a|g_a\rangle\langle g_a|\;,\qquad
\rho'_D=\sum_{a}p_b|h_b\rangle\langle h_b|\;.
\end{align}
These have von Neumann entropies
\begin{align}
&S'(A,B,D)=H(A)+\sum_ap_aS(B|a)\;,\\
&S'(A,C)=S(A)\;,\qquad S'(B,D)=S(B)\;,\\
&S'(A,B)=H(A,B)\;,\qquad S'(A)=S'(C)=H(A)\;,\qquad S'(B)=S'(D)=H(B)\;.
\end{align}

The proof now comes in a rush.  Recalling Eq.~(\ref{eq:HBgivenAbound}), we use the above results to write
\begin{align}
[S(A:B)-H(A:B)]&-[H_{\{E_a\}}(B|A)-S(B|A)]\nonumber\\
&=-S'(A,B,D)-S'(B)+S'(A,B)+S'(B,D)\nonumber\\
&=S'(A|B)-S'(A|B,D)\ge0\;.
\label{eq:WPMdiscord}
\end{align}
The inequality is precisely the expression of strong subadditivity for systems $A$, $B$, and $D$.  It shows that
\begin{equation}
\cM_{1b}^{{\rm(WPM)}}\ge\cM_{2c}^{{\rm(discord)}}=\sD(A\rightarrow B)\;.
\end{equation}
Concentrating now on discord, we write
\begin{align}
H_{\{E_a\}}(B|A)-S(B|A)
&=-S'(A,B,C,D)-S'(A)+S'(A,B,D)+S'(A,C)\nonumber\\
&=S'(B,D|A)-S'(B,D|A,C)\ge0\;,
\label{eq:discordnonneg}
\end{align}
where again the inequality is strong subadditivity, this time for systems $BD$, $A$, and $C$.  This inequality shows that discord is always nonnegative.

The equality conditions for strong subadditivity~\cite{Hayden2004a} can be applied to inequalities~(\ref{eq:WPMdiscord}) and~(\ref{eq:discordnonneg}).  From the latter inequality, we learn that $\cM_{2c}^{{\rm(discord)}}=\sD(A\rightarrow B)$ is zero if and only if $\rho_{AB}$ is diagonal in a conditional product basis pointing from $A$ to $B$. Datta~\cite{datta2010a}, in the proof of his Theorem~2, has shown how to use the equality conditions for strong subadditivity to obtain this necessary and sufficient condition for zero discord. From~(\ref{eq:WPMdiscord}), we learn that $\cM_{1b}^{{\rm(WPM)}}=\cM_{2c}^{{\rm(discord)}}=\sD(A\rightarrow B)$ if and only $\rho_{AB}$ is diagonal in a conditional product basis pointing from $B$ to $A$.  Combining these two results, we have that the WPM measure is zero if and only if $\rho_{AB}$ is diagonal in a product basis.

\section{Projective measurements vs.~POVMs for WPM and discord}
\label{app:projvsPOVM}

In this Appendix, we elaborate an example given by WPM~\cite{wu09a}, which exhibits a joint state that requires the use of rank-one POVMs, not just orthogonal-projection-valued measurements, to maximize the classical mutual information in evaluating the WPM measure.  We extend these results to show that for the same states, rank-one POVMs are required for evaluating the discord.

Consider a joint state
\begin{equation}\label{eq:appjoint}
\rho_{AB}=\sum_{j=1}^{d_B}p_j\rho_j\otimes|e_j\rangle\langle e_j|=
\sum_{j=1}^{d_B}p_j\rho_j\otimes P_j\;,
\end{equation}
where the states $|e_j\rangle$ make up an orthonormal basis for system~$B$.  The marginal states are given by
\begin{equation}
\rho_A=\sum_jp_j\rho_j\;,
\qquad
\rho_B=\sum_jp_j P_j\;,
\end{equation}
and this gives
\begin{align}
S(B)&=H(p_j)\;,\\
S(A,B)&=H(p_j)+\sum_jp_jS(\rho_j)\;,\\
S(A|B)&=S(A,B)-S(B)=\sum_jp_jS(\rho_j)\;,\\
S(B|A)&=S(A,B)-S(A)=H(p_j)+\sum_jp_jS(\rho_j)-S(A)\;,\\
S(A:B)&=S(B)-S(B|A)=S(A)-\sum_jp_jS(\rho_j)\;.
\end{align}
The quantum mutual information is the Holevo quantity for the ensemble of states $\rho_j$ with probabilities $p_j$.  This is not surprising because $\rho_{AB}$ describes a situation where $B$ sends a message to $A$: the message has the letters~$j$, with probabilities $p_j$; $B$ keeps a record
of the message in the orthogonal states $|e_j\rangle$ and encodes
the letters in the states $\rho_j$.

The state~(\ref{eq:appjoint}) has zero discord when communication goes from $B$ to $A$, i.e., $\sD(B\rightarrow A)=0$, because $\rho_{AB}$ is diagonal in a conditional product basis pointing from $B$ to $A$.  Generally, it has nonzero discord, $\sD(A\rightarrow B)$, when communication goes from $A$ to $B$.  The results of Appendix~\ref{app:WPMdiscord} show that the WPM measure equals $\sD(A\rightarrow B)$ for such states.  We return to discord below; for now, we focus on the WPM measure.  Given any unconditioned, local measurements on $A$ and $B$, the joint probability for results $a$ and $b$ is
\begin{equation}
p_{ab}=\sum_{j=1}^{d_B}p_jp_{a|j}p_{b|j}\;,
\end{equation}
where $p_{a|j}=\tr(E_a\rho_j)$ and $p_{b|j}=\tr(F_b P_j)$.  We can think of $p_{ab}$ as the marginal of a joint distribution for $a$, $b$, and~$j$:
\begin{equation}
p_{abj}=p_jp_{a|j}p_{b|j}\;.
\end{equation}
That $A$ and $B$ are conditionally independent means that
$p_{a|bj}=p_{a|j}$, which implies that $H(A|J)=H(A|B,J)$.  Thus we
have
\begin{equation}
H(A:J)-H(A:B)=H(A|B)-H(A|J)=H(A|B)-H(A|B,J)\ge0\;,
\end{equation}
where the final inequality follows from classical strong subadditivity, which says that additional conditioning cannot increase the entropy.  Measuring $B$ in the eigenbasis $|e_j\rangle$ gives $H(A:B)=H(A:J)$, so we can conclude that the maximum mutual information is attained on this measurement.  The WPM measure reduces to a form that only requires a maximization over the measurement on~$A$:
\begin{equation}
\cM_{1b}^{{\rm(WPM)}}
=S(A:B)-\max_{{\rm(b)}}H(A:B)=S(A:B)-\max_{\{E_a\}}H(A:J)\;.
\end{equation}

We now proceed to specialize the joint state $\rho_{AB}$ in four ways.  First, we assume that system $A$ is a qubit and that the ensemble states,
\begin{equation}\label{eq:rhoj}
\rho_j=\frac{1}{2}(I_A+\bm{\sigma\cdot n}_j)\;,
\end{equation}
are pure; i.e., the vectors $\bm{n}_j$ are unit vectors.  The measurement on $A$ is described by rank-one POVM elements
\begin{equation}\label{eq:purepovm}
E_a=q_a(I_A+\bm{\sigma\cdot m}_a)\;,
\end{equation}
where the vectors $\bm{m}_a$ are unit vectors.  The completeness relation for the POVM implies that the quantities $q_a$ make up a normalized probability distribution and that
\begin{equation}\label{eq:qmconstraint}
\sum_aq_a\bm{m}_a=0\;.
\end{equation}
The probability for result $a$, given state $\rho_j$, is
\begin{equation}
p_{a|j}=\tr(E_a\rho_j)=q_a(1+\bm{n}_j\bm{\cdot m}_a)\;,
\end{equation}
and the joint probability for results $a$ and $j$ is
\begin{equation}
p_{aj}=\tr(\rho_{AB}E_a\otimes P_j)=p_j\tr(E_a\rho_j)=
p_jq_a(1+\bm{n}_j\bm{\cdot m}_a)\;.
\end{equation}

The second specialization is to assume that the states $\rho_j$ are distributed so that
\begin{equation}\label{eq:symmetry}
\sum_{j=1}^{d_B}p_j\bm{n}_j=0
\qquad\Longleftrightarrow\qquad
\rho_A=\sum_{j=1}^{d_B}p_j\rho_j=\frac{1}{2}I_A\;.
\end{equation}
With this assumption we have that the probability for result~$a$ is
$p_a=q_a$ and thus that $S(B|A)=H(p_j)-S(A)=H(p_j)-1$
and $S(A:B)=S(A)=1$.  The classical mutual information is
\begin{align}
H(A:J)=H(A)-H(A|J)
=\sum_{j,a}p_{aj}\log(p_{a|j}/p_a)
=\sum_aq_aF(\bm{m}_a)\;,
\end{align}
where we define the function
\begin{equation}
F(\bm{m})\equiv
\sum_{j=1}^{d_B}p_j(1+\bm{n}_j\bm{\cdot m})\log(1+\bm{n}_j\bm{\cdot m})\;.
\end{equation}
The WPM measure is now given by $\cM_{1b}^{{\rm(WPM)}}=1-\widetilde H(A:J)$, where
\begin{equation}
\widetilde H(A:J)\equiv\max_{\{E_a\}}H(A:J)=
\max_{\{q_a,\bm{m}_a\}}\sum_a q_aF(\bm{m}_a)\;.
\end{equation}

Before going on to the third specialization, let's consider the quantum discord when conditioning on $A$.  We again make the first two specializations: a joint state of the form~(\ref{eq:appjoint}), with $A$ being a qubit, and the states $\rho_j$ being the pure states~(\ref{eq:rhoj}), distributed according to Eq.~(\ref{eq:symmetry}).  We measure the POVM~(\ref{eq:purepovm}) on $A$.  The probability for result $a$ is $p_a=q_a$, and the state of $B$, conditioned on result~$a$, is
\begin{equation}
\rho_{B|a}=\frac{\tr_A(E_a\rho_{AB})}{p_a}= \sum_{j=1}^{d_B} \frac{p_jp_{a|j}}{q_a} P_j= \sum_{j=1}^{d_B}p_j(1+\bm{n}_j\bm{\cdot m}_a) P_j\;,
\end{equation}
which has quantum entropy
\begin{equation}
S(B|a)=-\sum_{j=1}^{d_B}p_j(1+\bm{n}_j\bm{\cdot m}_a)
\log\bigl(p_j(1+\bm{n}_j\bm{\cdot m}_a)\bigr)\;.
\end{equation}
The conditional classical entropy that goes into the definition~(\ref{eq:tildeHBgivenA}) of discord becomes
\begin{equation}\label{eq:tildeHBgivenAapp}
\widetilde H(B|A)=\min_{\{E_a\}}\sum_a p_a S(B|a)
=H(p_j)-\max_{\{q_a,\bm{m}_a\}}\sum_a q_a F(\bm{m}_a)
=H(p_j)-\widetilde H(A:J)\;,
\end{equation}
so the discord is given by
\begin{equation}
\sD(A\rightarrow B)=\cM_{2c}^{{\rm(discord)}}
=\widetilde H(B|A)-S(B|A)
=1-\widetilde H(A:J)=\cM_{1b}^{{\rm(WPM)}}\;.
\end{equation}
As expected, the discord is the same as the WPM measure for this set of joint states.

Our third specialization is to assume that the ensemble probabilities are all equal, i.e., $p_j=1/d_B$, and the fourth, needed to work out examples,   is that the vectors $\bm{n}_j$ are symmetrically distributed, pointing to the vertices of an equilateral triangle or of a regular polyhedron.  Within this configuration, we first maximize the function $F(\bm{m})$.  The high degree of symmetry guarantees that the extrema of $F(\bm{m})$ occur along the symmetry axes of the triangle or polyhedron.  Having determined the maximum value, $F_{{\rm max}}$, one knows that this maximum is an upper bound on $\widetilde H(A:J)$.  The high degree of symmetry further guarantees that one can make up a POVM out of the directions $\bm{m}_a$ that give the maximum value, with $q_a$ chosen to be the same for all these directions; since this choice achieves the bound, one has $\widetilde H(A:J)=F_{{\rm max}}$.  Moreover, if no two of the directions $\bm{m}_a$ are oppositely directed, the upper bound cannot be achieved with a projection-valued measurement.  One ends up knowing, first, the common value of the WPM measure and quantum discord and, second, that the optimal measurement cannot be described by orthogonal projection operators.

For a triangle (tetrahedron) of states, the maximum value of $F$ is attained on the vectors that are directed opposite to the vectors that define the state.  The maximum value is $F_{{\rm max}}=\log\frac{3}{2}$ for the triangle and $F_{{\rm max}}=\log \frac{4}{3}$ for the tetrahedron.  The optimal measurement is the trine (tetrahedron) measurement that uses the triangle (tetrahedron) dual to the state triangle (tetrahedron).  Both the WPM measure and discord are equal to $1-\log \frac{3}{2}=\log \frac{4}{3}=0.415$ for the triangle and to $1-\log \frac{4}{3}=\log \frac{3}{2}=0.585$ for the tetrahedron.

We stress that the examples in this Appendix require that subsystem~$B$ have three or more Hilbert-space dimensions.  These examples thus do not exclude the possibility that projection-valued-measures suffice for WPM and discord for a pair of qubits.

It is worth noting that if one asks about demon discord, the quantity one needs to evaluate, instead of being the classical conditional entropy~(\ref{eq:tildeHBgivenAapp}), is the classical joint entropy~(\ref{eq:tildeHAB}).  Making the first two specializations gives a demon discord
\begin{equation}
\cM_{3c}^{{\rm(dd)}}
=\min_{\{q_a,\bm{m}_a\}}\biggl(H(q_a)-\sum_a q_a F(\bm{m}_a)\biggr)\;.
\end{equation}
The additional contribution from $H(q_a)$ prejudices this minimum toward using a smaller number of outcomes for the measurement on~$A$ and thus toward orthogonal-projection-valued measurements.  Indeed, for the triangle of states, with a trine measurement made in the dual triangle, the argument of the minimum is equal to 1.  For an orthogonal-projection-valued measurement, with $\bm{m}_1$ pointing toward one vertex of the state triangle and $\bm{m}_2$ in the opposite direction, the argument is equal to $\frac{4}{3}-\frac{1}{2}\log3=0.541$, which thus becomes the demon discord.  For this joint state, the optimal measurement for demon discord is orthogonal-projection-valued.

\section{Demon-based measures and rank-one POVMs}
\label{app:demons}

In this Appendix, we modify the formula for the net classical work to show that the local demons cannot do worse in terms of net classical work by restricting themselves to rank-one POVMs.

We use the general measurement formalism of Sec.~\ref{sec:localmeasurements}, which allows us to do strategies~(b) and~(c) simultaneously.  The state of system~$A$ after a measurement yields result~$a$ is
\begin{equation}
\rho_{A|a}=\frac{\sA_{a}(\rho_A)}{p_a}\;,
\end{equation}
and the state of system~$B$ after measurements that yield outcomes~$a$ and $b$ is given by
\begin{equation}
\rho_{B|ab}=\frac{\sB_{b|c(a)}(\rho_{B|a})}{p_{b|a}}\;.
\end{equation}
As the systems are transformed to the maximally mixed state, the local demons can extract work
\begin{equation}
W^+=\log(d_Ad_B)-\sum_ap_aS(\rho_{A|a})
-\sum_{a,b}p_{ab}S(\rho_{B|ab})\;.
\end{equation}
The cost of erasing the measurement record, given communication between the demons, is $W^-=H(A,B)$, giving a net classical work
\begin{equation}\label{eq:Wcapp}
W_c=W^+-W^-=\log(d_Ad_B)-H(A,B)-\sum_ap_aS(\rho_{A|a})
-\sum_{a,b}p_{ab}S(\rho_{B|ab})\;.
\end{equation}
The measure of nonclassical correlations requires maximizing $W_c$ over all possible measurements.

After finishing the first round of measurements, instead of extracting work, the local demons can make further measurements in the eigenbases of $\rho_{A|a}$ and $\rho_{B|ab}$.  The overall measurement is now described by rank-one POVMs.  After these measurements, the subsystems are left in pure states, so the local demons can extract work $W^+=\log(d_Ad_B)$ as the systems are transformed to the maximally mixed state, but they have a more detailed measurement record, so their erasure cost is greater.  If we let $\lambda_{\alpha|a}$ be the eigenvalues of $\rho_{A|a}$ and $\lambda_{\beta|ab}$ be the eigenvalues of $\rho_{B|ab}$, then after these measurements, the new erasure cost is
\begin{equation}
W^-=H(A,B)+\sum_ap_aH(\lambda_{\alpha|a})+\sum_{a,b}p_{ab}H(\lambda_{\beta|ab})\;.
\end{equation}
Since the classical entropies of the eigenvalues are the same as the quantum entropies, the net classical work is the same as that given in Eq.~(\ref{eq:Wcapp}).  The reduction in classical work from not using rank-one POVMs has been transferred to an increased erasure cost when making measurements described by rank-one POVMs.  We conclude that the demons cannot do worse by restricting themselves to rank-one POVMs, thus justifying our assumption of rank-one POVMs in Sec.~\ref{sec:Measures}.

\bibliography{dedicated}

\end{document}